\documentclass[pra,aps,onecolumn,preprint,superscriptaddress,amsmath,amssymb,floatfix]{revtex4}
\usepackage[breaklinks=true,colorlinks=true,linkcolor=blue,urlcolor=blue,citecolor=blue]{hyperref}
\usepackage{graphicx,url,float,dcolumn,bm,xcolor}
\usepackage{rotating}
\usepackage{pdflscape}
\usepackage[caption = false]{subfig}
\usepackage{soul}
\def\prn#1{{\left(#1\right)}}
\def\cbrk#1{{\left\{#1\right\}}}

\def\abrk#1{{\langle#1\rangle}}

\def\ts#1{{_{\mbox{\scriptsize #1}}}}

\def\mb{\mathbf}
\def\bs{\boldsymbol}

\usepackage{textcomp}

\def\prn#1{{\left(#1\right)}}
\def\cbrk#1{{\left\{#1\right\}}}

\def\abrk#1{{\langle#1\rangle}}

\begin{document}

\title{Characterization of the Global Network of Optical Magnetometers to search for Exotic Physics (GNOME)}

\author{S.~Afach}
\affiliation{Helmholtz Institut Mainz, Johannes Gutenberg-Universit\"{a}t, 55099 Mainz, Germany}

\author{D.~Budker}
\affiliation{Helmholtz Institut Mainz, Johannes Gutenberg-Universit\"{a}t, 55099 Mainz, Germany}
\affiliation{Department of Physics, University of California, Berkeley, California 94720-7300, USA}
\affiliation{Lawrence Berkeley National Laboratory, Berkeley, CA 94720}

\author{G.~DeCamp}
\affiliation{Department of Physics, California State University -- East Bay, Hayward, California 94542-3084, USA}

\author{V.~Dumont}
\affiliation{Department of Physics, University of California, Berkeley, California 94720-7300, USA}

\author{Z.~D.~Gruji\'c}
\affiliation{Physics Department, University of Fribourg, Chemin du Mus\'{e}e 3, CH-1700 Fribourg, Switzerland}

\author{H.~Guo}
\affiliation{Center for Quantum Information Technology, Peking University, Beijing 100871, PR China}

\author{D.~F.~Jackson~Kimball}
\affiliation{Department of Physics, California State University -- East Bay, Hayward, California 94542-3084, USA}

\author{T.~W.~Kornack}
\affiliation{Twinleaf LLC, 300 Deer Creek Drive, Plainsboro, NJ 08536, USA}

\author{V.~Lebedev}
\affiliation{Physics Department, University of Fribourg, Chemin du Mus\'{e}e 3, CH-1700 Fribourg, Switzerland}

\author{W.~Li}
\affiliation{Center for Quantum Information Technology, Peking University, Beijing 100871, PR China}

\author{H.~Masia-Roig}
\affiliation{Helmholtz Institut Mainz, Johannes Gutenberg-Universit\"{a}t, 55099 Mainz, Germany}

\author{S.~Nix}
\affiliation{Department of Physics and Astronomy, Oberlin College, Oberlin, Ohio 44074, USA}

\author{M.~Padniuk}
\affiliation{Institute of Physics, Jagiellonian University, prof. Stanis\l{}awa
\L{}ojasiewicza 11, 30-348, Krak\'{o}w, Poland}

\author{C.~A.~Palm}
\affiliation{Department of Physics, California State University -- East Bay, Hayward, California 94542-3084, USA}

\author{C.~Pankow}
\affiliation{Center for Interdisciplinary Exploration and Research in Astrophysics (CIERA) and Department of Physics and Astronomy,
Northwestern University, Evanston, IL 60208, USA}

\author{A. Penaflor}
\affiliation{Department of Physics, California State University -- East Bay, Hayward, California 94542-3084, USA}

\author{X.~Peng}
\affiliation{Center for Quantum Information Technology, Peking University, Beijing 100871, PR China}

\author{S.~Pustelny}
\affiliation{Institute of Physics, Jagiellonian University, prof. Stanis\l{}awa
\L{}ojasiewicza 11, 30-348, Krak\'{o}w, Poland}

\author{T.~Scholtes}
\affiliation{Physics Department, University of Fribourg, Chemin du Mus\'{e}e 3, CH-1700 Fribourg, Switzerland}

\author{J.~A.~Smiga}
\affiliation{Helmholtz Institut Mainz, Johannes Gutenberg-Universit\"{a}t, 55099 Mainz, Germany}


\author{J.~E.~Stalnaker}
\affiliation{Department of Physics and Astronomy, Oberlin College, Oberlin, Ohio 44074, USA}

\author{A.~Weis}
\affiliation{Physics Department, University of Fribourg, Chemin du Mus\'{e}e 3, CH-1700 Fribourg, Switzerland}

\author{A.~Wickenbrock}
\affiliation{Helmholtz Institut Mainz, Johannes Gutenberg-Universit\"{a}t, 55099 Mainz, Germany}

\author{D.~Wurm}
\affiliation{Technische Universit\"at M\"unchen, 85748 Garching, Germany}

\date{\today}

\begin{abstract}
The Global Network of Optical Magnetometers to search for Exotic physics (GNOME) is a network of geographically separated, time-synchronized, optically pumped atomic magnetometers that is being used to search for correlated transient signals heralding exotic physics. The GNOME is sensitive to nuclear- and electron-spin couplings to exotic fields from astrophysical sources such as compact dark-matter objects (for example, axion stars and domain walls). Properties of the GNOME sensors such as sensitivity, bandwidth, and noise characteristics are studied in the present work, and features of the network's operation (e.g., data acquisition, format, storage, and diagnostics) are described. Characterization of the GNOME is a key prerequisite to searches for and identification of exotic physics signatures.
\end{abstract}

\maketitle

\setlength{\parindent}{5ex}

\section{Introduction}
\label{Sec:Introduction}

There are a number of recent and ongoing experiments using atomic magnetometers \cite{2013:budker} to search for exotic fields mediating spin-dependent interactions \cite{Saf18RMP}. The basic concept of such experiments is to search for anomalous energy shifts of Zeeman sublevels caused by exotic fields rather than ordinary electromagnetic fields. For example, there are experiments searching for exotic spin-dependent interactions constant in time as evidence of new long-range monopole-dipole \cite{1992:venema,2008:heckel,2017:kimball} and dipole-dipole interactions \cite{Hun13}, where the Earth is the source of mass or polarized electrons, respectively. There are also experiments searching for shorter-range exotic spin-dependent interactions using local sources that can be modulated, such as laboratory-scale masses or polarized spin samples \cite{You96,Chu13,Vas09,Lee18}. A number of experiments test local Lorentz invariance (LLI) by moving a comagnetometer with respect to a hypothetical background field, either via a rotatable platform for the experiment \cite{Smi11} or through the motion of the Earth itself relative to this background field \cite{Gem10}.

The {\textbf{G}}lobal {\textbf{N}}etwork of {\textbf{O}}ptical {\textbf{M}}agnetometers to search for {\textbf{E}}xotic physics (GNOME) collaboration is searching for an entirely different class of effects: signals from {\emph{transient events}} \cite{Pos13,Kim18AxionStars,budker2015data} that could arise from an exotic field of astrophysical origin passing through the Earth during a finite time.  While a single magnetometer system could detect such transient events, it would be exceedingly difficult to confidently distinguish a true signal generated by exotic physics from false positives induced by occasional abrupt changes of magnetometer operational conditions (e.g., magnetic-field spikes, laser mode hops, electronic noise, etc.).  Effective vetoing of prosaic transient events (false positives) requires an array of individual, spatially distributed magnetometers to eliminate spurious local effects.  Furthermore, a global distribution of sensors is beneficial for event characterization, providing, for example, the ability to resolve the velocity of the exotic field by observing the relative timing of transient events at different sensors \cite{Pos13}.  The Laser Interferometer Gravitational Wave Observatory (LIGO) collaboration has developed sophisticated data analysis techniques \cite{And01,All12,klimenko2016method} to search for correlated transient signals using a worldwide network of gravitational-wave detectors. Recently, the GNOME collaboration demonstrated that these and similar analysis techniques can be applied to data from synchronized magnetometers \cite{2013:pustelny}.

The sensitivity and bandwidth of each GNOME magnetometer is discussed below. Many existing sensors in the network have magnetometric sensitivities $\lesssim 100~{\rm fT/\sqrt{Hz}}$ and bandwidths $\approx 100~{\rm Hz}$, and with planned upgrades all magnetometers should be able to achieve these specifications. Each magnetometer is located within a multi-layer magnetic shield to reduce the influence of magnetic noise and perturbations, while retaining sensitivity to exotic fields and interactions \cite{Kim16}. Even with magnetic-shielding techniques, there is inevitably some level of magnetic field transients from both local sources as well as due to global effects (such as solar wind, changes of the Earth's magnetic field, etc.).  Therefore, each GNOME magnetometer uses auxiliary sensors (unshielded magnetometers, accelerometers, gyroscopes, and other devices) to measure relevant environmental conditions, allowing for exclusion/vetoing of data for which there are identifiable sources generating transient signals. These auxiliary sensors are monitored, and if their readings go beyond an acceptable range the data collected from that particular GNOME magnetometer are flagged as suspect during that time.

The signals from the GNOME magnetometers are recorded with accurate timing provided by the Global Positioning System (GPS) using a custom GPS-disciplined data acquisition system \cite{Wlo14}. Many of the current (and future) GNOME magnetometers have a temporal resolution of $\lesssim 10~{\rm ms}$ (determined by the magnetometers' bandwidths and data sampling rate), enabling resolution of events that propagate at the speed of light (or slower) across the Earth ($2R_E/c \approx 40~{\rm ms}$, where $R_E$ is the Earth's radius). Because of the broad geographical distribution of sensors, the GNOME acts as an exotic physics ``telescope'' with a baseline comparable to the Earth's diameter.

The initial scientific focus of the GNOME is a search for correlated transient signals generated by terrestrial encounters with massive compact dark-matter objects composed of axion-like particles (ALPs), such as ALP domain walls \cite{Pos13,2013:pustelny} and ALP stars \cite{Kim18AxionStars}. Based on the characteristic relative velocity between virialized dark matter objects and the solar system, the Earth would travel at $\sim 10^{-3}c$ (where $c$ is the speed of light) through the dark-matter object, leading to $\sim 40$~s delays between transient signals at different sites (depending on the geometry of the encounter, see discussion in Ref.~\cite{Rob17}). The GNOME is also sensitive, for example, to cosmic events generating a propagating wave burst of an exotic field \cite{Car94,Arv15}, or to long-range correlations produced by a fluctuating \cite{Ell04} or oscillating \cite{Bud14} exotic field whose time-averaged value is zero. The specific techniques and tools used to analyze GNOME data in order to search for each of these various exotic physics targets are somewhat different and will be described in detail in future publications. As previously noted, one example of such analysis, based on methods employed by the LIGO collaboration \cite{And01,All12}, is described in Ref.~\cite{2013:pustelny}. A closely related, complementary approach is being pursued by the GPS.DM collaboration to search for a different class of compact dark-matter objects using atomic clocks as sensors rather than atomic magnetometers \cite{Der14,Rob17} (a similar technique has been pursued in Refs.~\cite{wcislo2017experimental,Wcislo2018ojh}). There have also been recent proposals to search for transient signals generated by exotic physics using networks of interferometers \cite{Sta15,Sta16} and resonant bar detectors \cite{Jac15,Arv16,Bra17}. We envision that in the future GNOME data can be used in conjunction with data from clocks, interferometers, and resonant cavities to form a multi-sensor network to search for exotic physics signatures.

In this work, we present a discussion of the experimental techniques used to acquire the GNOME data and essential characteristics of those data. The active GNOME system during its first collective data acquisition period (Science Run 1, beginning June 6th, 2017 at 12AM UTC) consisted of six dedicated optical atomic magnetometers located at five geographically separated stations: Berkeley, California, USA (two sensors); Fribourg, Switzerland; Hayward, California, USA; Krakow, Poland; and Mainz, Germany (listed alphabetically by city). This article describes characteristics of the sensors comprising the GNOME system during Science Run 1. In particular, we discuss the magnetometer setups at each station, the auxiliary sensors used to veto false positives, and the computer infrastructure for acquisition, storage, and transfer of data.  We report on a series of test runs used to study the operation of the network infrastructure and to characterize the response and sensitivity of all existing GNOME stations. A second science run was carried out in December of 2017 with four additional optical atomic magnetometers in Beijing, China; Daejeon, South Korea; Hefei, China; and Lewisburg, Pennsylvania, USA participating. Characterization of the expanded GNOME (consisting of 10 magnetometers) used in Science Run 2 will be described in a future article. There are also a number of new GNOME stations planned or under construction (in Be'er Sheva, Israel; Berlin, Germany; Canberra, Australia; Jena, Germany; Los Angeles, California, USA; Oberlin, Ohio, USA; and Stuttgart, Germany). A third GNOME Science Run is presently underway, aiming for an extended ($\sim$ 1 year) observation period with more active GNOME stations. For Science Run 3, new standards for magnetometer calibration and performance checks have been adopted to improve data quality motivated by the results of the studies presented here.

\section{Experimental Setup of the Magnetometer Network}

\subsection{General Characteristics of the Magnetometers}

\begin{table*}[!ht]
  \centering
  \caption{Basic characteristics of GNOME magnetometers participating in Science Run 1. The direction of the leading field defines the sensitive axis of the magnetometer as well as the relative sign of transient signals with respect to other stations. The altitudes and azimuths of the leading field directions are given according to the local horizontal coordinate system, where altitude~$\in \cbrk{-90^\circ,90^\circ}$ ($90^\circ$ indicating the direction to the zenith) and azimuth~$\in \cbrk{0^\circ,360^\circ}$ ($0^\circ = 360^\circ$ indicating north). Note that if the altitude is $\pm 90^\circ$, the azimuth is undefined. In the row headings, GS HF stands for ``ground-state hyperfine.''}
  \label{Table:BasicCharacteristics}
  \begin{tabular}{lccccccc}
    \hline
    \hline
    \rule{0ex}{3.6ex} Property 						& Berkeley~1		& Berkeley 2		& Fribourg			& Hayward			& Krakow			& Mainz				\\
    \hline
    \rule{0ex}{3.6ex} Atomic species        		& $^{133}$Cs		& $^{133}$Cs		& $^{133}$Cs		& $^{85}$Rb			& $^{87}$Rb			& $^{87}$Rb			\\
    \rule{0ex}{3.6ex} GS HF level probed  			& $F=4$				& $F=4$				& $F=4$				& $F=3$  			& $F=2$				& $F=2$				\\
	\rule{0ex}{3.6ex} Leading field (nT)			& 489				& 1930				& 650				& 1495				& 1158				& 525   			\\
  \rule{0ex}{3.6ex} Larmor frequency (Hz)			& 1710				& 6756				& 2274				& 6975    			& 8100				& 3679				\\
    \rule{0ex}{3.6ex} Longitude	  	                & 122.2572$^\circ$W	& 122.2572$^\circ$W	&~~7.1575$^\circ$E	& 122.0540$^\circ$W &~19.9046$^\circ$E	&~~8.2346$^\circ$E	\\
    \rule{0ex}{3.6ex} Latitude	                    &~37.8722$^\circ$N	&~37.8722$^\circ$N	&~46.7930$^\circ$N	&~37.6564$^\circ$N 	&~50.0286$^\circ$N	&~49.9906$^\circ$N	\\
    \rule{0ex}{3.6ex} Leading field altitude 		& 0$^\circ$	     	& 90$^\circ$		& 0$^\circ$			&-90$^\circ$  		& 0$^\circ$			&-90$^\circ$		\\
    \rule{0ex}{3.6ex} Leading field azimuth			& 28$^\circ$ 		& -		    		&190$^\circ$		& -    				& 45$^\circ$		& -			\\
    \hline
    \hline
  \end{tabular}
\end{table*}

All the existing GNOME magnetometers are optically pumped atomic magnetometers (see Refs.~\cite{2013:budker,budker2007optical,Weis16} for reviews) that measure the spin-precession frequency of alkali atoms by observing the time-varying optical properties of the alkali vapor with a probe laser beam. The alkali vapor is contained within an antirelaxation-coated cell \cite{Ale02,Cas09,Bal10}, and the vapor cell is located inside a set of magnetic-field coils that enable control of homogeneous longitudinal and transverse components of an applied field $\mb{B}_0$ (leading field) as well as (for some stations) magnetic field gradients. The cell and coil system are mounted within a multi-layer magnetic shield that provides shielding of external fields to a part in $10^5$ or better. Some basic characteristics of the various GNOME magnetometer stations are listed in Table~\ref{Table:BasicCharacteristics}.

\subsection{Example of a magnetometer setup: Fribourg}
\label{Sec:FribourgSetup}

\begin{figure*}
\includegraphics[width=6.5 in]{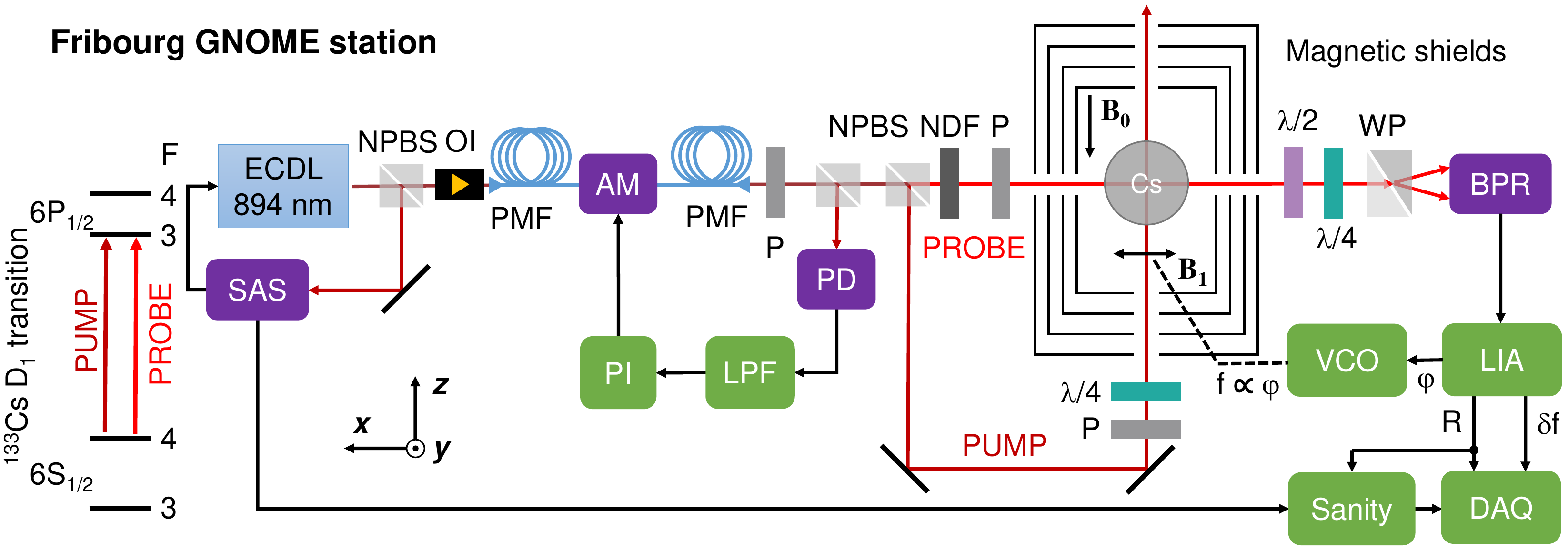}
\caption{Schematic diagram of the experimental setup for the Fribourg GNOME magnetometer. Left-hand side of figure shows the level scheme for the probed atoms, in this case the $^{133}$Cs $D_1$ transition is used both for pumping and probing. The applied fields $\mb{B}_0$ (along $-\hat{\mb{z}}$) and $\mb{B}_1$ along the $x$-axis are generated by coils within the magnetic shields. At the center of the magnetic shield system is a spherical antirelaxation-coated Cs vapor cell. The coordinate axes for the experiment referenced in the text are indicated by the arrows at the lower left. Red lines/arrows indicate laser paths, black lines/arrows indicate electronic connections. Notation: ECDL = extended-cavity diode laser; SAS = saturated absorption spectroscopy setup; NPBS = non-polarizing beamsplitter; OI = optical isolator; PMF = polarization-maintaining fiber; AM = amplitude modulator; PI = proportional-integral control loop electronics; LPF = low-pass filter; P = linear polarizer; PD = photodiode; NDF = neutral density filter; $\lambda/2$ = half-wave plate; $\lambda/4$ = quarter-wave plate; WP = Wollaston prism;  BPR = balanced photo receiver; VCO = voltage controlled oscillator; LIA = lock-in amplifier; Sanity = sensors and electronics to check data quality (Sec.~\ref{Sec:Sanity}); DAQ = GPS-disciplined data acquisition system (Sec.~\ref{Sec:GPS-DAQ}). The frequency $f$ of $\mb{B}_1$ is tuned by a phase-locked loop in which the phase $\varphi$ delivered by the LIA is used to control a VCO. The magnitude $R$ of the measured signal read from the LIA is one of the parameters used by the sanity monitor to confirm the system is operating properly and that the data are reliable. Both $R$ and the deviation $\delta f$ of $f$ from its initial setting are inputs to the GPS DAQ.}
\label{Fig:ExptSetupFribourg}
\end{figure*}

A typical example of the experimental setup of the optical atomic magnetometers comprising the GNOME is the Fribourg magnetometer shown in Fig.~\ref{Fig:ExptSetupFribourg}. Descriptions of the experimental setups for the other GNOME magnetometers are given in Appendix~\ref{Appendix:ExptSetups}. The Fribourg GNOME magnetometer is an rf-driven magnetometer in pump-probe geometry with circular-dichroism detection. It is located in a temperature-controlled container cabin on the roof of the Physics Department building at the University of Fribourg (Switzerland). The system consists of an optical table setup, incorporating the magnetometer within a magnetic shield, and an electronics and laser rack.

The optical table is a 35~mm thick 60$\times$40~cm$^2$ breadboard, lying on vibration-isolating layers of foam and Sorbothan, mounted on a 15~mm thick aluminum plate resting on four passively air-damped supports (Thorlabs, model PTH602).
The whole setup is enclosed in a 7.5~cm thick custom-made styrofoam box (60$\times$90$\times$65~cm$^3$) for passive thermal isolation.

A solid state diode laser (Toptica, model DL100 pro) is used to generate the laser light used for optical pumping and probing of the Cs atomic sample. The laser frequency is stabilized to the  $F{=}4{\rightarrow}F'{=}3$ hyperfine component of the Cs D1 transition (894~nm) by a custom-made saturated absorption spectroscopy (SAS) unit employing laser current modulation at 195~kHz.
The stabilization circuit feeds the transimpedance-amplified (TEM, model PDA-S) photodiode (Thorlabs, model DET36A/M) signal to a digital lock-in amplifier, followed by two PID controllers (Toptica, model Digilock 110) that adjust the laser diode injection current and the external cavity piezo voltage. The Toptica laser system allows for relocking of the laser frequency using a remote desktop application.
The power transmitted through an auxiliary Cs reference cell (room temperature, buffer-gas-free, uncoated -- not shown in the schematic of Fig.~\ref{Fig:ExptSetupFribourg}) is detected with a photodiode (Thorlabs, model DET36A/M), whose  signal level is checked  by the sanity system (Sec.~\ref{Sec:Sanity}). The fluorescence light emitted by the Cs atoms within the reference cell is monitored with an infrared-sensitive camera whose readout is accessible via the internet, allowing remote monitoring and control of the lock status.

The main laser beam passes through an optical isolator and is coupled into a polarization-maintaining fiber (PMF, Thorlabs, model P3-780PM-FC), which carries the light to the magnetometer proper. The PMF is connected to an integrated electro-optical amplitude modulator (EOM; Jenoptik, model AM905). A small fraction of the light (split off by a non-polarizing beamsplitter after the fiber) is recorded with a photodiode (Thorlabs, model DET36A/M) and used in an active proportional and integral (PI) feedback system stabilizing the power at the photodiode. The servo-loop contains a low-pass filter (cutoff frequency $\sim$1~Hz), so that only long-term laser-power and polarization drifts are corrected.

The heart of the setup is a custom-made evacuated paraffin-coated (28~mm diameter) Cs vapor cell \cite{Cas09}, placed in the central inner volume of a 4-layer mu-metal magnetic shield system (Twinleaf, model MS-1).
The static (leading) magnetic field ($B_0$=650~nT) along $-\hat{\mb{z}}$ is produced with the Twinleaf coil system inside the innermost shield that is driven with a thermally insulated Magnicon (model CSE-1) current source (coils and current source not shown in Fig.~\ref{Fig:ExptSetupFribourg}).

The mount of the vapor cell holds a pair of RF Helmholtz coils (each with 3 turns of 5~cm diameter wire loops), producing an oscillating magnetic field $B_1$ of $2.65~\mathrm{nT_{rms}}$ along $x$, and a Helmholtz-like coil (two single 6.5~cm diameter loops), that can produce a magnetic field along $\mathbf{B}_0$ for calibration purposes (Sec.~\ref{ssec:bandwidth}).

The fiber-coupled light is collimated to a beam diameter of $\sim$2~mm and split into a pump beam and a probe beam using a non-polarizing beam splitter. The pump beam ($\sim$150~\textmu W) is circularly-polarized with a linear polarizer and quarter-wave plate and propagates along $\hat{\mb{z}}$. The linearly-polarized probe beam ($\sim$30~\textmu W) is guided to traverse the cell along $-\hat{\mb{x}}$, orthogonal to the pump beam's propagation.

The transmitted probe beam passes through a half-wave-plate and a quarter-wave-plate, after which it is split by a Wollaston prism (Thorlabs, model WP10) into two orthogonal linearly-polarized components.
The axes of the Wollaston prism are set at $\pm 45^\circ$ with respect to the incoming probe beam's polarization.
The difference between the powers of the two components is detected with a balanced photoreceiver (Thorlabs, model PDB210A).
We note that all other GNOME stations record the rotation of the transmitted probe beam's plane of polarization generated by the spin-polarized medium's circular birefringence (CB), which is related to the different indices of refraction for $\sigma_+$ and $\sigma_-$ light.
The Fribourg station, on the other hand, detects the probe beam's ellipticity that is induced by  the vapor's circular dichroism (CD), due to the different absorption coefficients for $\sigma_+$ and $\sigma_-$ light.
Compared to the CB detection, which requires a frequency-detuned probe beam,  CD detection is most efficient with resonant light.
In this way, both the pump and the probe beam can be derived from the same laser, which eases operation and reduces the cost of the set-up.

A 19"-rack (mounted on a rigid baseplate with vibration-isolating Sorbothan feet) holds the laser system, the magnetometer read-out electronics and a personal computer (PC) for experiment control and data-streaming.
%
%
The magnetometer signal from the balanced photoreceiver is analyzed with a digital lock-in amplifier (LIA; Zurich Instruments, model HF2LI).
The phase $\varphi$ of the oscillatory signal from the balanced photoreceiver with respect to the $B_\mathrm{1}(t)$ oscillation at $\omega_\mathrm{RF}$ has the typical $\arctan$ dependence on the detuning $\delta\omega{=}\omega_\mathrm{RF}{-}\omega_\mathrm{L}$ from the Larmor frequency $\omega_\mathrm{L}$.
The linear $\varphi{\propto}\delta\omega$ dependence near  $\delta\omega{\approx}0$ is used as an error signal for generating the RF frequency $\omega_\mathrm{RF}{\propto}\varphi{-}\varphi_\mathrm{0}$, using a voltage-controlled oscillator (VCO).
We refer to this mode of operation, in which the phase is actively stabilized to a given (geometry-dependent \cite{Weis16}) reference value $\varphi_\mathrm{0}$ as phase-locked operation.
This phase-locked-loop (PLL) oscillates at the instantaneous Larmor frequency $\omega_\mathrm{L}$, that it follows with a -3dB bandwidth of $\sim$100~Hz.

The HF2LI features an analog voltage output (conversion  factor of 0.286~V/Hz) that represents the deviation of the actual oscillation frequency from a preset frequency, that is fed to the GPS DAQ box (Sec.~\ref{Sec:GPS-DAQ}). In order to avoid aliasing effects during the analog-digital conversion a -48~dB/octave roll-off Butterworth low-pass filter (SRS, model SIM965) with a 170~Hz cut-off frequency efficiently suppresses frequency components above the Nyquist frequency (250~Hz) of the DAQ box's ADC converter.

The magnetometer's $R$-signal (the square root of the quadrature sum of the in-phase and out-of-phase LIA outputs) is output as a scaled voltage (conversion factor of 10~V/V$_\mathrm{rms}$) and fed to a second GPS DAQ box channel as well as to the sanity system, as a check to ensure that the magnetic resonance condition is fulfilled and the amplitude is above a set threshold. Similarly to the other stations, a dedicated GPS DAQ box channel receives the output signal of the sanity system (Sec.~\ref{Sec:Sanity}), which flags data that auxiliary measurements indicate not to be reliable. The GPS DAQ box is connected to the data-streaming PC which uploads data to the central GNOME server in Mainz, Germany (Sec.~\ref{Sec:DataTransfer}).

\subsection{GPS-disciplined data acquisition system}
\label{Sec:GPS-DAQ}

An important aspect of the operation of the GNOME is synchronous measurement of magnetometer readouts between the various stations spread all over the Earth. This requires precise global timing, which needs to be available across the Earth. Currently, the only source fulfilling these requirements is the global-positioning system (GPS). Depending on the number of visible satellites, the system can provide signals with time-accuracy of better than 50~ns.

To take full advantage of the GPS timing, the Krakow GNOME group designed and built a dedicated GPS-disciplined data acquisition system (DM Technologies Data Acquisition System), which provides the ability to store several analog signals with precise timing and read a few digital sensors with less accuracy. While the system was carefully described elsewhere \cite{Wlo14}, here we recall its most important features.

The GNOME GPS DAQ box is a stand-alone data acquisition system. The heart of the system is an AMR7-core Atmel microcontroller clocked by a 48-MHz quartz oscillator, which is responsible for handling time reference, controlling and synchronizing data acquisition, storing data to a memory card, and enabling communication with a computer. Additionally, the microcontroller handles communication with a user via outputs to a liquid-crystal display and control buttons mounted in the front panel of the device.

In our system, a time reference is provided by a GPS time receiver (Trimble Resolution T). This module, being an integral part of the acquisition system, is connected with a GPS antenna and, if enough satellites are visible (more than 3), provides a pulse-per-second (PPS) signal with an accuracy of 45~ns (at the 3-$\sigma$ level). The signal is transmitted to the microcontroller, which handles it with the highest priority and initiates data acquisition (opens analog-to-digital converters). After the pulse, the time receiver also transmits additional information such as the number of visible satellites, antenna position (longitude, latitude, and altitude), temperature, and any reported warning. This information is stored by the box for reference.

The acquisition system used in the experiments discussed in the present work has four analog input channels enabling measurement of signals in four bipolar ranges ($\pm 1.25$~V, $\pm 2.5$~V, $\pm5$~V, and $\pm 10$~V) and sampling rates of 1~S/s, 2~S/s, 4~S/s, $\dots$ , 512~S/s, 1024~S/s. The system implements a special software algorithm, that provides uniformly distributed samples (see Ref.~\cite{Wlo14} for more details). This ensures that even in case of a drift of the 48-MHz clock, the samples are measured in equal intervals.

The acquired data are stored in a memory card (a solid-state disk, SD card) in 1-minute text files (FAT32 file structure). Since typically each SD card ensures only a finite number of storage cycles (between 10,000 and 100,000), the card used in our system uses a special wear-leveling algorithm (equal usage of disk space) enabling a higher number of storage cycles (1,000,000). Application of the card enables buffering of the data and provides means for independent operation of the system (a 4-GB card offers 24 hours of independent operation), but after that period the data are overwritten. The storage capabilities of the box may be improved by replacing the card with a larger capacity card. Due to the FAT32 file structure system, the box accepts up to 32-GB SD cards.

The system communicates with a local computer over a universal serial bus (USB) connection. When it is connected with the computer, it appears as the device (GPS DAQ) at a specific serial port. Typically, the acquisition system transmits stored data to the computer, but it can also be put into a special service mode, enabling maintenance of the time receiver module. To avoid data overwriting, the oldest data are first transmitted from the box to the PC. Particularly, after reestablishment of data transmission after a communication failure between the local computer and the GPS DAQ box, the oldest data are transmitted first while new data are continuously stored. Depending on the size of buffered data and number of acquired channels, this process may take between a few minutes to several hours to complete.

While GPS time is provided with a $\pm$45~ns precision, it is not the only source of delay and uncertainty present in the system. Particularly, transmission of signals from the antenna (typically situated on the roof of a bulding in which the GNOME station is located) introduces some hundreds of ns of delay (200~ns with a 50-m cable). Less than 100~ns of delay is introduced by conversion of the analog signal to its digital form. The largest source of delay, however, is introduced by the software (entering into the acquisition mode after the timing signal arrives), which spans between 1.4~\textmu s to 1.6~\textmu s. The cumulative delay of the system corresponds to roughly 2~\textmu s with an uncertainty of about $\pm$200~ns. In case of signals propagating at the speed of light, this corresponds to a position uncertainty of less than 100~m; for compact dark matter objects with relative velocities with respect to the Earth's frame of $\sim 10^{-3}c$, based on the timing accuracy the corresponding position uncertainty is less than 0.1~m.

\subsection{Sanity Monitor}
\label{Sec:Sanity}

As mentioned in the introduction, an important aspect of background noise reduction in GNOME is identifying data that may exhibit transients because of measurable environmental factors or technical issues rather than exotic physics. To address this, the Fribourg GNOME group has developed an automated system to check for the ``sanity'' of the data. A digital output is sent to the GNOME GPS DAQ indicating whether or not the data is ``sane'' (i.e., free of any environmental perturbations detected above-threshold). ``Insane'' data from a magnetometer can then be flagged and ignored at the analysis stage. This way, known errors affecting individual stations can be detected and dealt with, thereby reducing both background noise and false positive events. Presently the data is flagged as ``insane'' if any of the monitored signals exceed the user-determined thresholds.

Errors that might produce transient spikes in the data stream can be caused by, for example, loss of laser lock or failure of a system component, mechanical shocks (e.g., due to earthquakes --- an effect that has actually been observed on several occasions by the Fribourg GNOME station, for example), magnetic or electric pulses from neighboring devices, or human activity. If such identifiable errors are not properly marked in the data they might be falsely interpreted as evidence of exotic physics.

The GNOME sanity monitor system is based on an Arduino MEGA 2560 microcontroller board \cite{ardui} which features ADC/DAC channels and is additionally equipped with the Arduino 9 axes motion shield using the Bosch BNO055 SiP (system in package) that integrates a triaxial 14-bit accelerometer, a triaxial 16-bit gyroscope with a range of ±2000 degrees per second and a triaxial geomagnetic sensor \cite{bosch} on a single chip. The sensors are read out in 40~ms intervals. The nine readings of all three integrated sensors are compared to a rolling average of the last 31 consecutive points. If the deviation is larger than a user-defined bound, the microcontroller triggers an ``insanity'' event. The standard deviation of the rolling average of 31 consecutive measurements emerging due to statistical noise of the sensors is typically $\lesssim 0.1\,\mathrm{^{\circ}/s}$ for the gyroscope, $\lesssim 0.03\,\mathrm{m^2/s}$ for the accelerometer and $\lesssim 0.8$\,\textmu T for the magnetometer axial components.

A box housing the microcontroller is mounted on the optical table of the GNOME station near the magnetic shielding. In addition to the integrated sensors, the sanity monitor features several analog input channels arranged on a separate break-out box that can be used to monitor critical system parameters of the particular GNOME station, such as the error signal of the laser lock(s), the magnetometer signal amplitude, readings of temperature sensors connected to different parts of the setup, or devices monitoring the beam position(s) and so on. Digital input pins are used for interlock mechanisms (e.g. checking, for example, that if the system is enclosed in a thermally insulating box, the box's lid is closed) and for a manual override switch that can be used in case the operator has to make changes to the experimental setup, enabling the station to remain on-line and continuously streaming data during tests and maintenance. If one of the monitored channels falls out of its ``sane'' range, which is specified using the dedicated sanity software, the sanity monitor will indicate an ``insane" state to the GPS DAQ, marking the data to be rejected at the analysis stage.

A dedicated Python-based software communicates with the Arduino microcontroller using the built-in USB interface. The graphical user interface (GUI) of the software enables the user to set the number of channels and the respective ``sane'' ranges of the input channels of the system and to program the configuration to the microcontroller memory. After setting up the sanity monitor, the microcontroller can run without being connected to the PC. However, if the PC connection is kept in place, the software is able to monitor the actual states of the channels and provides detailed logging of the acquired signals. In case of an ``insane'' event, a log-entry will be written, storing the detailed state of all the input channels and integrated sensors, thus allowing to trace the origin of the sanity state failure in a post-analysis.

An important issue that will be studied in detail in the near future is the correlation between glitches in the magnetometer data indicating a transient signal above noise and the output of the sanity monitor. This study will analyze the degree to which the sanity monitor is successful at reducing the rate of single-detector false-positives and determine optimum settings for thresholds. In the present investigation, each GNOME station determined independently sanity monitor settings and thresholds at appropriate levels to veto severe malfunctions (for example, the laser system losing lock).

\subsection{Data format, transfer, and storage}
\label{Sec:DataTransfer}

After the GPS DAQ digitizes the data, the data are transferred to a computer and a Python-based program parses and saves the data in files using the Hierarchical Data Format HDF5 \cite{hdf5}. Each GNOME HDF5 file contains 60 seconds of data. HDF5 provides a tree data structure with multi-dimensional datasets, where each dataset is an $N$-dimensional table. Datasets also include meta-data and header-data referred to as ``Attributes.'' The data types in HDF5 datasets can be as simple as a single data type per dataset, or more complex data-structures (where, for example, every entry can have multiple data types, i.e., a combination of integers, floating-point numbers, etc.). The data storage in HDF5 uses the low-level binary format of the data in question. For example, floating-point numbers are stored using the Institute of Electrical and Electronics Engineers (IEEE) 754 standard \cite{ieee754}, which was chosen for its efficient data compression in order to save diskspace \cite{zlib}.

To ensure that it is possible to systematically process the saved GNOME data, a storage standard for the files was developed, referred to as the GNOME Data Standard (GDS). The HDF5-compatible GDS is continuously developed with backward-compatibility in mind. As an example of the rules of the GDS, the GDS dictates that every valid GNOME data file must contain a dataset denoted the ``default dataset'' which contains the primary signal related to the measured effective magnetic field at a station and a sanity signal. The GDS restricts the format of the default dataset to one or two dimensions, and demands that the default dataset contains mandatory attributes, such as the sample rate of the data, the time, and others, including an attribute with an equation in string form that is used to convert the raw data (e.g., voltage) into magnetic field units (e.g., pT). The GDS is inclusive, not exclusive in nature; meaning that it does not restrict stations from adding any additional data to the data files, but dictates that certain data elements must be included in the data files. This enables storage of station-specific data for additional analysis.

After locally writing data to HDF5 files, the data are uploaded to a data-server located in Mainz and maintained by the Mainz GNOME group. The data transfer is done through a server/client pair developed with C++. The client is available to all stations and has a GUI that works across multiple platforms (Windows, Linux, and Mac). The server is a terminal program that runs on a Linux server, and receives multiple connections from multiple clients. A client, upon connecting to the server, is set to monitor a single directory. The directory is expected to be filled with data in HDF5 format that complies with the GDS. When a new data file appears, it is added to a queue for uploading, and is tested for its integrity and compliance to the GDS. If the data are compliant with GDS, its Message Digest (MD5) checksum is calculated and data packets are created and sent to the server with TCP/IP wrapped in SSL (Secure Sockets Layer) encrypted packets. The server checks the data integrity through the MD5 (Message Digest 5) algorithm and compliance with GDS, and finally saves the data in a redundant data storage that is maintained by the Helmholtz Institut at Gutenberg Universit\"at in Mainz.

\section{Sensitivity of the network to exotic fields}
\label{sec:exotic-field-sensitivity}

Although the GNOME is a network of magnetometers, ultimately the goal of the GNOME is not to search for magnetic-field transients but rather to search for exotic fields coupling to atomic spins. While it is practical to measure and compare the sensitivities of the magnetometers in units of magnetic field, depending on the nature of the exotic physics searched for these sensitivities must be re-scaled by various factors.

The types of exotic fields searched for by the GNOME can essentially be described as pseudo-magnetic effective fields $\mathcal{B}\ts{eff}$ causing energy shifts of Zeeman sublevels and, equivalently, torques on atomic spins. Unlike true magnetic fields, however, the couplings of an exotic field $\bs{\Upsilon}$ to various particles' spins are not generally proportional to their magnetic moments \cite{Saf18RMP}. For example, in some axion models there is relatively strong coupling of the axion field to proton spins, weak coupling to neutron spins, and no coupling to electron spins \cite{Kim79,Shi80}. This is important for interpretation of data from the GNOME, as the expected response of each sensor to $\bs{\Upsilon}$ needs to be appropriately scaled depending on the assumed nature of the interaction. As an example, one could imagine two magnetometers utilizing different atomic species with similar magnetic moments but nuclei having valence protons spins pointing along or opposite to the total atomic angular momentum vector, respectively (for example, this occurs in $^{85}$Rb and $^{87}$Rb \cite{2017:kimball}). In this case, a magnetic field transient would generate similar responses in the two magnetometers, but an exotic field transient that coupled primarily to proton spins would generate signals of opposite signs in the two magnetometers. If the proper scaling was not taken into account at the analysis stage, such an event might be regarded as a false positive and vetoed.

To interpret data from the GNOME, we employ a relatively simple framework for modeling the response of magnetometers to exotic spin-dependent interactions (reviewed in Refs.~\cite{Saf18RMP} and \cite{Kim15}), valid to first-order for electrons and valence nucleons, based on the Russell-Saunders approximation for the atomic structure and the Schmidt model for the nuclear structure. Table~\ref{Table:spin-content} shows the relevant intrinsic factors related to the magnetometers' sensitivities to exotic fields: the Land\'e $g$-factors, projection of the electron spin polarization along the total atomic angular momentum direction normalized to the total atomic angular momentum
\begin{align}
\sigma_e = \frac{\abrk{\mb{S}_e \cdot \mb{F}}}{F(F+1)}~,
\label{Eq:electron-fractional-polarization}
\end{align}
and the normalized projection of the proton spin polarization along the total atomic angular momentum direction
\begin{align}
\sigma_p = \frac{\abrk{\mb{S}_p \cdot \mb{F}}}{F(F+1)}~,
\label{Eq:proton-fractional-polarization}
\end{align}
where $\mb{S}_e$ is the electron spin, $\mb{F}$ is the total atomic angular momentum, and $\abrk{\cdots}$ denotes the expectation value of the considered quantity. Table~\ref{Table:spin-content} also lists the sensitivity scaling factors $\sigma_e/g_F$ and $\sigma_p/g_F$. All magnetometers used in the present GNOME are based on alkali atoms whose nuclei have valence protons and are thus primarily sensitive to proton as opposed to neutron spin interactions. Given a coupling strength $\kappa_i$ of $\bs{\Upsilon}$ to particle $i$ ($i=e,p$ for electron or proton in this case), the field $\mathcal{B}\ts{eff}$ measured by a magnetometer based on a particular alkali atom in a given ground-state hyperfine level is:
\begin{align}
\mathcal{B}\ts{eff} = \prn{\frac{\kappa_e\sigma_e + \kappa_p\sigma_p}{g_F}} \bs{\Upsilon}~.
\label{Eq:ObservedField}
\end{align}
This scaling will be accounted for in analysis of GNOME data. Relative signs and amplitudes of observed transient signals should be consistent with a single value for the coupling constant $\kappa_i$ of each standard model fermion for all magnetometers. Note that since $g_F \approx 2\sigma_e$, the scaling factor for electrons ($\approx 1/2$) is the same for all species and ground states to first order.

\begin{table}
\caption{Comparison of the normalized projection of the electron/proton spin along the total atomic angular momentum vector [Eqs.~\eqref{Eq:electron-fractional-polarization} and \eqref{Eq:proton-fractional-polarization}] and Land\'e $g$-factors for $^{85}$Rb, $^{87}$Rb, and $^{133}$Cs; see Ref.~\cite{Kim15} for more details.}
\medskip \begin{tabular}{lccccc} \hline \hline
\rule{0ex}{3.6ex} Atom~(state)~~~~ & ~~~$\sigma_e$~~~ &  ~~~$\sigma_p$~~~ & ~~~$g_F$~~~ & ~~~$\sigma_e/g_F$~~~ & ~~~$\sigma_p/g_F$~~~ \\
\hline
\rule{0ex}{3.6ex} $^{85}$Rb~($F=3$) & 0.17 & -0.12 & 0.33 & 0.50 & -0.36 \\
\rule{0ex}{3.6ex} $^{85}$Rb~($F=2$) & -0.17 & -0.17 & -0.33 & 0.50 & 0.50 \\
\rule{0ex}{3.6ex} $^{87}$Rb~($F=2$) & 0.25 & 0.25 & 0.50 & 0.50 & 0.50 \\
\rule{0ex}{3.6ex} $^{87}$Rb~($F=1$) & -0.25 & 0.42 & -0.50 & 0.50 & -0.84 \\
\rule{0ex}{3.6ex} $^{133}$Cs~($F=4$) & 0.13 & -0.10 & 0.25 & 0.50 & -0.40 \\
\rule{0ex}{3.6ex} $^{133}$Cs~($F=3$) & -0.13 & -0.12 & -0.25 & 0.50 & 0.48 \\
\hline \hline
\end{tabular}
\label{Table:spin-content}
\end{table}

\begin{figure*}
\includegraphics[scale=1]{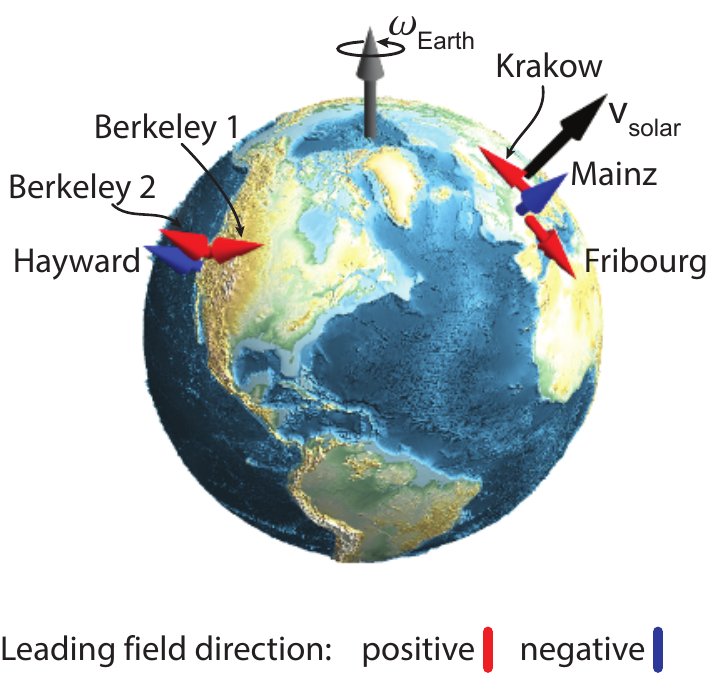}
\caption{Illustration of the locations and directions of the leading fields of the GNOME magnetometers listed in Table~\ref{Table:BasicCharacteristics}; red arrows indicate fields along that direction, blue arrows indicate fields oriented oppositely to that direction. Also shown for reference is $\mb{v}\ts{solar}$, the relative velocity of the solar system with respect to the galactic rest frame, at a particular time (relative to the Earth frame, this vector changes direction due to the motion of the Earth). $\mb{v}\ts{solar}$ is the most probable relative velocity between the Earth and a compact dark matter object \cite{Rob17}, and also the most probable axis along which $\mathcal{B}\ts{eff}$ is directed in a number of models of such dark matter objects \cite{Pos13,Kim18AxionStars}.}
\label{Fig:Globe}
\end{figure*}

Another important feature of the network that must be accounted for in data analysis is the fact that all GNOME magnetometers used in Science Run 1 employed a leading field $\mb{B}_0$ of hundreds of nT applied along the directions listed in Table~\ref{Table:BasicCharacteristics}. Therefore the magnetometers are first-order sensitive to exotic transient fields $\mathcal{B}\ts{eff}$ parallel to $\mb{B}_0$ and only second-order sensitive to fields $\mathcal{B}\ts{eff}$ orthogonal to $\mb{B}_0$. The sensitivity to $\mathcal{B}\ts{eff}$ thus varies from magnetometer-to-magnetometer based on the orientation of $\mb{B}_0$ with respect to the galactic rest frame, and this sensitivity varies daily due to Earth's rotation. The locations of the magnetometers and the various directions of these leading fields, as well as the relative velocity of the solar system with respect to the galactic rest frame ($\mb{v}\ts{solar}$), are illustrated in Fig.~\ref{Fig:Globe}.

\begin{figure*}
\includegraphics[scale=1]{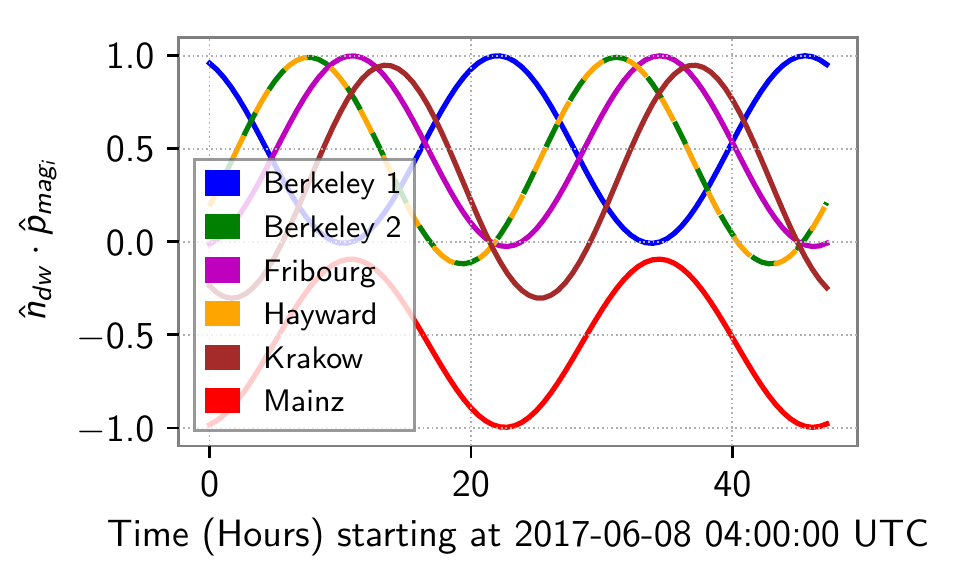}
\caption{Daily modulation of the effective sensitivity of the different vector magnetometers comprising the GNOME due to the rotation of the Earth. In this calculation it is assumed that the effective field $\mathcal{B}\ts{eff}$ is along $\mb{v}\ts{solar}$. In this case, the effective sensitivity is scaled by the dot product between the unit vector along the leading field $\mb{B}_0$ ($\hat{\mb{p}}\ts{mag,i}$) for each magnetometer and the unit vector along $\mb{v}\ts{solar}$ ($\hat{\mb{n}}_{dw}$), except for the case of Hayward where in the data the positive direction of the field was defined to be opposite to the direction of the leading field (this has been changed for future science runs to be consistent with other stations).}
\label{Fig:DirSens}
\end{figure*}

The daily modulation of the sensitivity of the various magnetometers is illustrated in Fig.~\ref{Fig:DirSens}, where it is assumed that $\mathcal{B}\ts{eff}$ is oriented along $\mb{v}\ts{solar}$ (determined in this calculation by the direction from the Earth's center to the star Deneb in the Cygnus constellation, towards which the Sun moves). The dot product between the unit vector along the leading field $\mb{B}_0$ ($\hat{\mb{p}}\ts{mag,i}$) for each magnetometer and the unit vector along $\mb{v}\ts{solar}$ ($\hat{\mb{n}}_{dw}$) is shown for each sensor. This factor can be dealt with in a variety of ways in the analysis stage, and in general the data analysis will not assume a particular direction of $\mathcal{B}\ts{eff}$ but rather scan over possible directions or leave the direction as a free parameter.

\section{Magnetometer characterization}
\label{Sec:MagnetometerCharacterizationData}

\subsection{Bandwidth measurements}
\label{ssec:bandwidth}

\begin{figure*}
\includegraphics[scale=1.0]{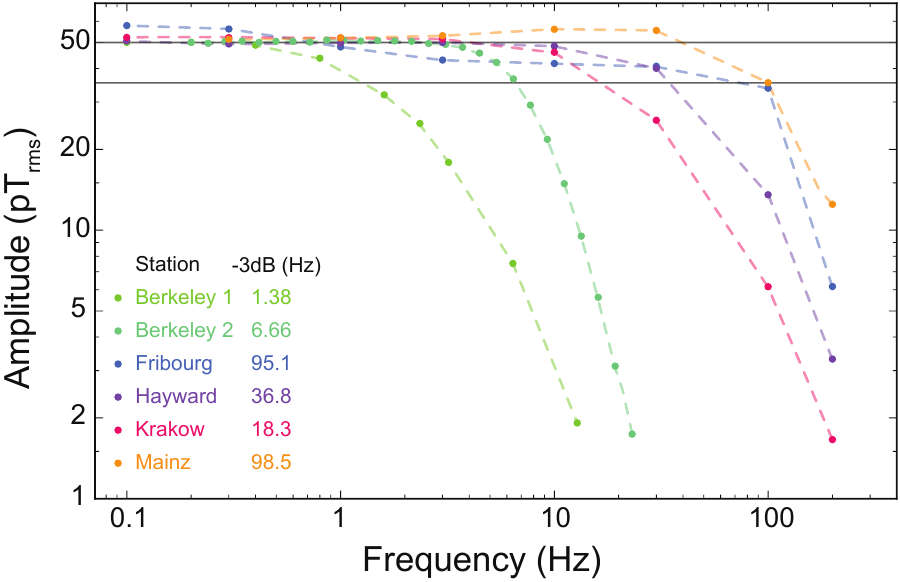}
\caption{Bandwidth measurements (dots) for all stations carried out during a GNOME test run on 2 November 2016, except the Berkeley data which were recorded during a later independent experiment. Data points are connected by lines to guide the eye. The Berkeley data were re-scaled to 50~pT/rms at low frequencies (see text). The -3dB points $\Delta f_\mathrm{-3dB}$ (shown in the legend) have been inferred from the intersection point of the lines with the horizontal line at $50/\sqrt{2}~\mathrm{pT_{rms}}$.}
\label{fig:bandwidth}
\end{figure*}

Determining the bandwidths of the constituent GNOME magnetometers is crucial for interpretation of the data in terms of transient exotic-physics signals. Since each GNOME magnetometer has a finite bandwidth, the GNOME has a frequency-dependent sensitivity to both magnetic fields and exotic pseudo-magnetic fields that couple to atomic spins. Thus in order to interpret the observation of a correlated transient signal and/or derive constraints on exotic physics from a null measurement, magnetometer bandwidths must be taken into account to relate the detected signals to the actual fields producing the signals. In order to determine the bandwidths of the GNOME magnetometers, a coordinated calibration of all stations was carried out during a 24-hour test run starting on 2 November 2016; the results of these measurements are shown in Fig.~\ref{fig:bandwidth}. For a given time window ($180\,\mathrm{s}$) the participating stations used dedicated coils internal to the shields to synchronously apply an oscillating calibration magnetic field of amplitude $B\ts{cal}=50~\mathrm{pT_{rms}}$ aligned parallel to the static, leading magnetic field. The oscillation frequency of $B\ts{cal}$ was consecutively shifted to higher values ($0.1,\,0.3,\,1,\,3,\,10,\,30,\,100,\,200\,\mathrm{Hz}$) in subsequent time windows.

Figure~\ref{fig:bandwidth} shows the recorded rms-values of the measured magnetometer responses, determined by fits of sine-wave functions to the streamed data from each of the stations. (Note that data shown for the Berkeley stations were obtained independently from the coordinated calibration run on 2 November 2016 using a different, equivalent methodology: application of an additional small modulation of frequency $\nu\ts{mod}$ to the AOM input signal and recording of the magnetometer signal amplitude demodulated at $\nu\ts{mod}$.)

In terms of data analysis, an important conclusion to be drawn from these bandwidth measurements is that while transient events varying on characteristic time scales of $\sim 1$~s could produce signals in all studied magnetometers, signals from transient events varying on time scales faster than $\sim 1$~s would be relatively suppressed in some magnetometers, increasingly so as characteristic time scales become shorter. Thus in the data analysis algorithms the effective sensitivity of the GNOME to transient signals must be appropriately scaled based on these bandwidth measurements. The bandwidths of free-running magnetometers (such as the Hayward station) are determined by the characteristic relaxation rate of the atomic spin polarization, as this sets the time scale over which the contribution of atomic spin polarization evolution under the influence of the fields is effectively averaged to produce the measured optical rotation signal (see discussions in, for example, Refs.~\cite{2013:budker,Weis16,budker2002resonant}).

Since, in principle, the spin-precession (Larmor) frequency responds instantaneously to changes in the field, various techniques, such as the implementation of phase-locked loops (PLLs), can be used to increase the magnetometer bandwidths (as is done, for example, in the Fribourg station, Sec.~\ref{Sec:FribourgSetup}). For Science Run 3 one of the performance standards for all GNOME magnetometers is to have calibrated bandwidths of $\approx 100~{\rm Hz}$, which is achieved by implementing appropriate PLLs (or similar techniques) at all stations.

\subsection{Time and frequency characteristics of raw data}
\label{ssec:TimeFrequency}

\begin{figure}
\includegraphics[width=\columnwidth]{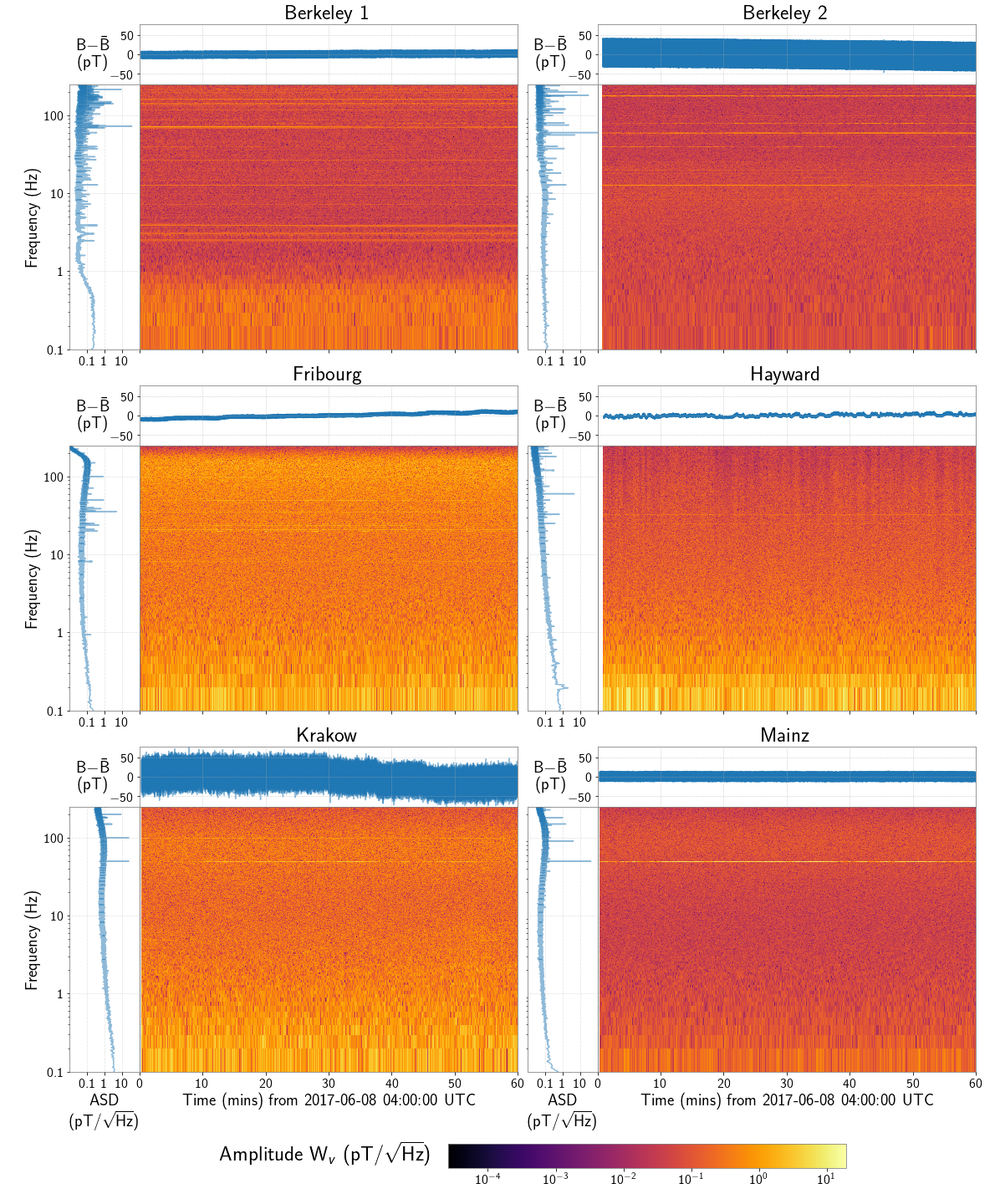}
\caption{One hour of characteristic data for each station (UTC 04:00:00 to 05:00:00 on 8 June 2017): the upper plots show the time series of the measured magnetic field, the left-hand plots show the magnetic field amplitude spectral density, and the central plots are spectrograms generated as described in the text.}
\label{spectrograms}
\end{figure}

Figure \ref{spectrograms} shows characteristic time series of the magnetometer signals and corresponding amplitude spectral densities and spectrograms \cite{scipy-spectrogram-note} for one hour (UTC 04:00:00 to 05:00:00 on 8 June 2017) of uninterrupted data from all six magnetometers. Each magnetometer's time series had its respective mean value subtracted and the magnetic fields are plotted on the same scale for every station to facilitate visual comparison. The amplitude spectral density for each station was computed by using the Welch method \cite{scipy-welch-method-note} where the data were divided into 10 overlapping segments. Individual amplitude spectral density curves are then computed for each segment and subsequently averaged.

To produce the spectrograms, the signals are divided into 10-second-duration segments, each containing 5,000 individual samples. Each segment overlaps with the previous segment by 2,500 samples. For each segment, a one-dimensional discrete Fourier transform is calculated using the Fast Fourier Transform (FFT) algorithm \cite{Cooley1965}. The absolute value of the Fourier-transformed data are normalized to the maximum signal during the one-hour acquisition period and then plotted.

The amplitude spectral density plots and spectrograms reveal several notable features that should be accounted for in the analysis of GNOME data. For example, a number of stations consistently observe relatively large signals at line frequencies (50 or 60 Hz) or harmonics of line frequencies. In several cases these line signals are observed even when they are well outside the bandwidths of the magnetometers (compare Figs.~\ref{fig:bandwidth} and \ref{spectrograms}), which suggests these signals originate from electronic interference rather than line signals leaking into the leading magnetic fields applied through the internal coils. Post-acquisition, digital notch filters at the line frequencies (as discussed in Secs.~\ref{sec:psd} and \ref{Sec:long-term}) or noise-whitening techniques \cite{2013:pustelny} can be applied to the data in order to reduce spurious effects related to the line signals. The Hayward, Fribourg, and both Berkeley stations also show noticeable, consistent signals at other frequencies. Efforts to identify and reduce/eliminate sources of apparent magnetometer noise are ongoing at every GNOME station.

\section{Power Spectral Density (PSD)}
\label{sec:psd}
\begin{figure*}[!h]
\includegraphics[width=0.85\textwidth]{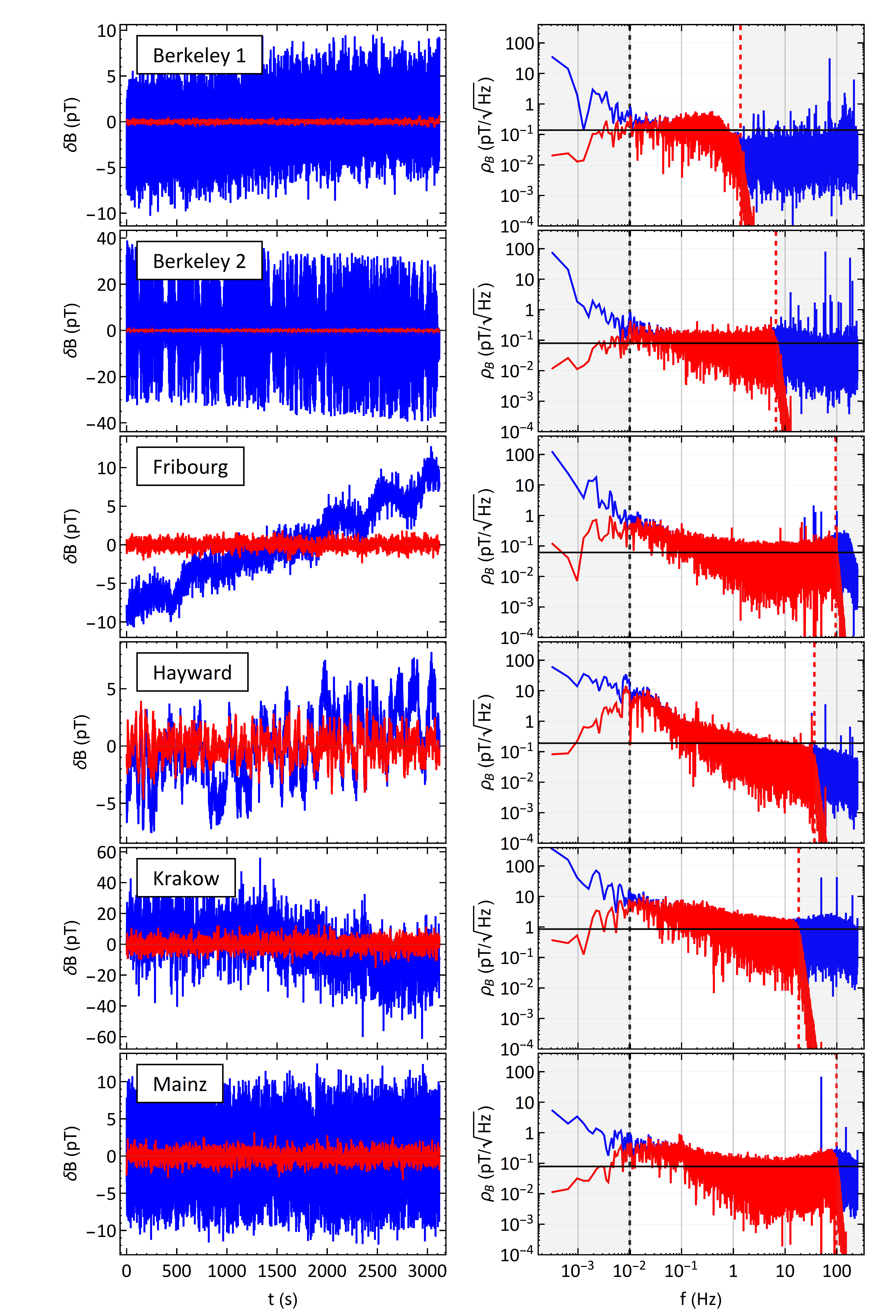}
\caption{Left: Time series of one hour of raw data (blue) superimposed on filtered data (red). Right: Square-root of the power spectral densities of the raw (blue) and filtered (red) data. The vertical dashed red lines indicate each magnetometer's $f_c$. The vertical dashed black lines represent the cut-on frequency of the high-pass filter. The horizontal black lines represent $\rho_B$ as inferred from the histograms in Sec.~\ref{sec:histo}. Note that the plots on the left-hand side have different vertical scales.
}
\label{Fig:FFTs}
\end{figure*}

The left column of Fig.~\ref{Fig:FFTs} shows the time series from one hour (UTC 04:00:00 to 05:00:00 on 8 June 2017) of uninterrupted data from all six magnetometers.
The blue lines represent the raw time series, and the overlaid red lines result from filtering the latter by a 6th order Butterworth amplitude low-pass filter (LPF) with an amplitude transfer function
\begin{equation}
\mathcal{T}=\frac{1}{\sqrt{1+\left(f/f_c\right)^{2n}}}\,\text{ with } n=6\,,
\label{eq:LPF}
\end{equation}
where $f_c$ is the -3dB cut-off frequency of each individual magnetometer response as listed in Fig.~\ref{Fig:FFTs}.
This low-pass filtering suppresses noise contributions at $f>f_c$, which are of electronic rather than magnetic origin.
We note that for an n-th order LPF, characterized by $f_c$, the bandwidth $\Delta f$ of a flat ($\mathcal{T}=1$ in $[ 0,\Delta f ]$, $\mathcal{T}=0$ elsewhere) filter transmitting the same power (of white noise) as the Butterworth LPF of Eq.~\eqref{eq:LPF} is given by
\begin{equation}
\Delta f = \sqrt{\frac{\pi }{2n\sin
   \left(\frac{\pi }{2 n}\right)}}\,f_c  \approx 1.006 \,f_c\,
\text{ for }
n=6\,.
\label{eq:BW}
\end{equation}
Since some of the stations feature large amplitude monochromatic oscillations, the latter were additionally removed by (4th order Butterworth) notch filters, centered at 24/36/39/50~Hz for Fribourg, and 50/90~Hz for Mainz, respectively.
We applied the notch filters only for spectra containing strong harmonic oscillations at frequencies below the corresponding $f_c$ cut-off.
In addition we applied a forward-backward (drift-removing) high-pass (1st order Butterworth) filter with a cut-on frequency of 10~mHz in both the increasing and decreasing time sequence, shown as dashed vertical line in Fig.~\ref{Fig:FFTs}.
%
%

The right column of Fig.~\ref{Fig:FFTs} shows the  square-root of the power spectral densities (rPSD) of the time series, all spectra being displayed with the same range of horizontal and vertical axes.
The Fourier transforms are calculated using the full 3600$\times$500 data points, and have been decimated for display purposes
The blue and red spectra represent the rPSD of the raw and filtered data, respectively.
Most magnetometers feature a reasonably flat spectrum down to (or below) 0.1~Hz.
The rise of the rPSD at lower frequencies reflects signal drifts.
%

%
%
\section{Allan Standard Deviation (ASD)}
\label{sec:asd}
\begin{figure*}
\includegraphics[width=\textwidth]{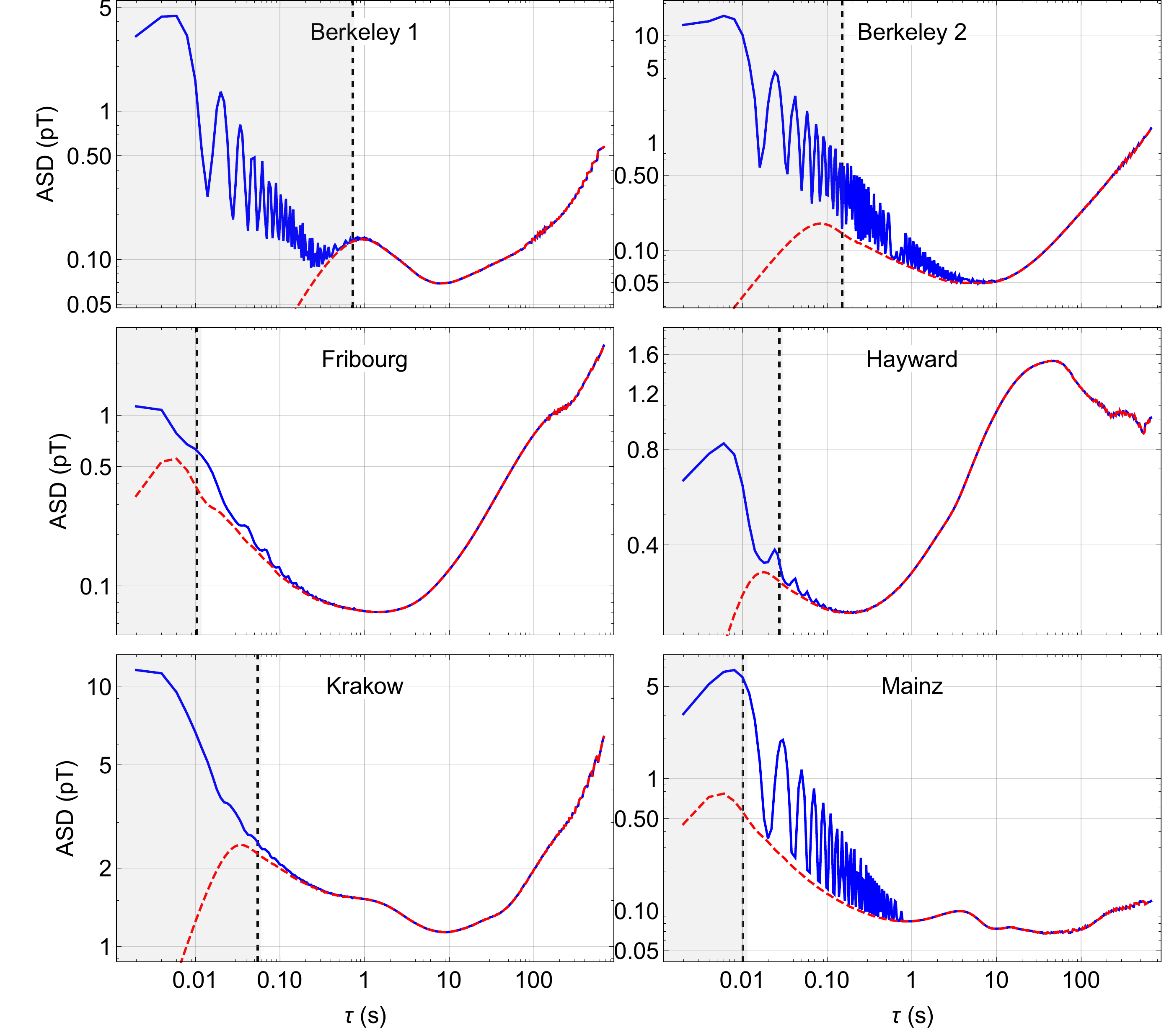}
\caption{Allan Standard Deviations (raw data in solid blue, LPF- and notch-filtered data in dashed red) of all six magnetometers, calculated from the same one hour data set as used in Fig.~\ref{Fig:FFTs}. The vertical dashed lines mark the inverse of each magnetometers bandwidth, so that data within the region marked in light gray are not relevant for the ASD analysis. Note the different vertical scale ranges of the individual graphs.}
\label{Fig:AllanDeviations}
\end{figure*}

\begin{figure*}
\includegraphics[width=0.8\textwidth]{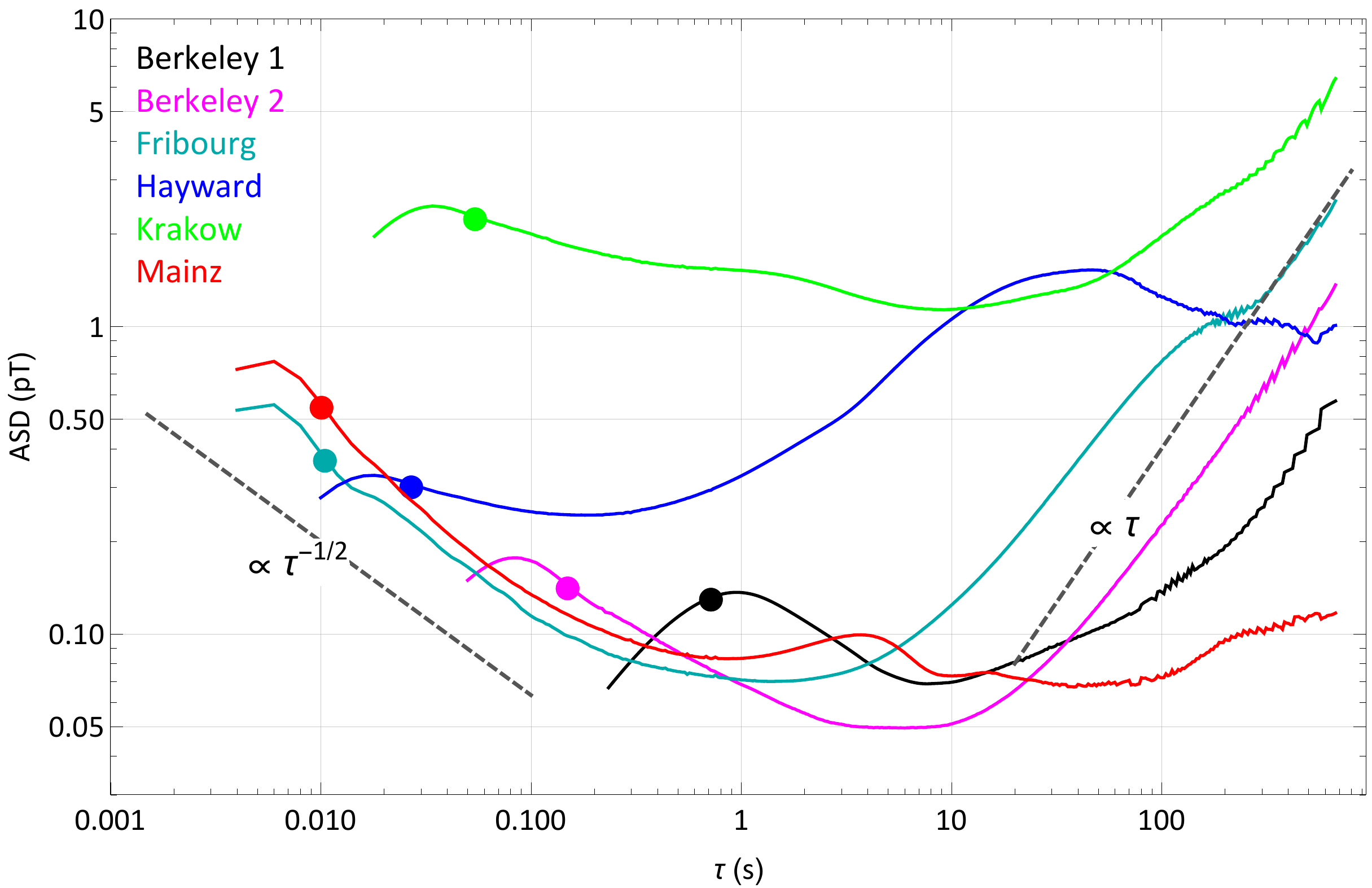}
\caption{Superposition of all filtered ASDs from Fig.~\ref{Fig:AllanDeviations}. The colored dots on each curve mark the inverse of each magnetometer's bandwidth. Black dashed lines indicate the limiting cases of pure white noise ($\propto\tau^{-1/2}$) and linear drift ($\propto\tau$).
}
\label{Fig:AllanDeviationsAll}
\end{figure*}
The goal of the GNOME is to search for exotic physics by detecting coincident transient events for which the magnetometer signal exceeds the background noise during some specific time interval.
The duration of reasonable time intervals depends critically on the noise and stability of each magnetometer.
In practice, one wants to compare time-averaged signals in data bins of duration $T\ts{bin}$ to the background magnetometer signals calculated by rolling averages over times $T\ts{bgd} \gg T\ts{bin}$. Such averaging would, depending on the noise characteristics, be optimized for detection of transient signals of duration $T\ts{bin}$.
%
%

While the PSD analysis is most useful for investigating the system behavior at short time scales, the appropriate tool for characterizing the long-term behavior is the so-called Allan Standard Deviation (ASD) \cite{Allan66}.
Figure~\ref{Fig:AllanDeviations} shows the ASDs of all stations, calculated from the same data set used in Fig.~\ref{Fig:FFTs}.
The ASDs from the raw data are shown as blue lines.
The vertical dashed lines represent the inverse of each magnetometer`s cut-off frequency $f_c$, so that data in the range marked in light gray carry no relevant information.
Some of the magnetometers show strong monochromatic oscillations, which make the interpretation of the ASD plots difficult.
For this reason, we show in the same graphs also the data after notch- and LPF-filtering as red dashed lines.

Since the vertical scales of the individual graphs in Fig.~\ref{Fig:AllanDeviations} are all different, we superpose all filtered ASDs in Fig.~\ref{Fig:AllanDeviationsAll}, together with guide lines indicating the slopes of the typically encountered ASD$(\tau)\propto\tau^{-1/2}$ and $\propto\tau$ slopes.
As long as the ASD decreases with increasing $\tau$ one can reduce the statistical uncertainty on the background reference level by increasing the time $\tau$ over which the signal is averaged.
The optimal integration time $T\ts{bgd}$ is thus $\tau_\ts{min}$, i.e., the time for which the ASD reaches a minimum.
In this respect the Fribourg and Mainz magnetometers show an optimal performance in the region from 10~ms to a few seconds, while Berkeley 2 reaches down to 100~ms only.
The other magnetometers' performances are limited either by their reduced bandwidth (Berkeley 1) or their reduced sensitivity (Hayward and Krakow).
Since the time when the data discussed here were taken, the performance of most magnetometer stations has been improved and shall be described in follow-up publications discussing searches for exotic physics.
%

%
\section{Histograms}
\label{sec:histo}
\begin{figure*}
\includegraphics[width=0.7\textwidth]{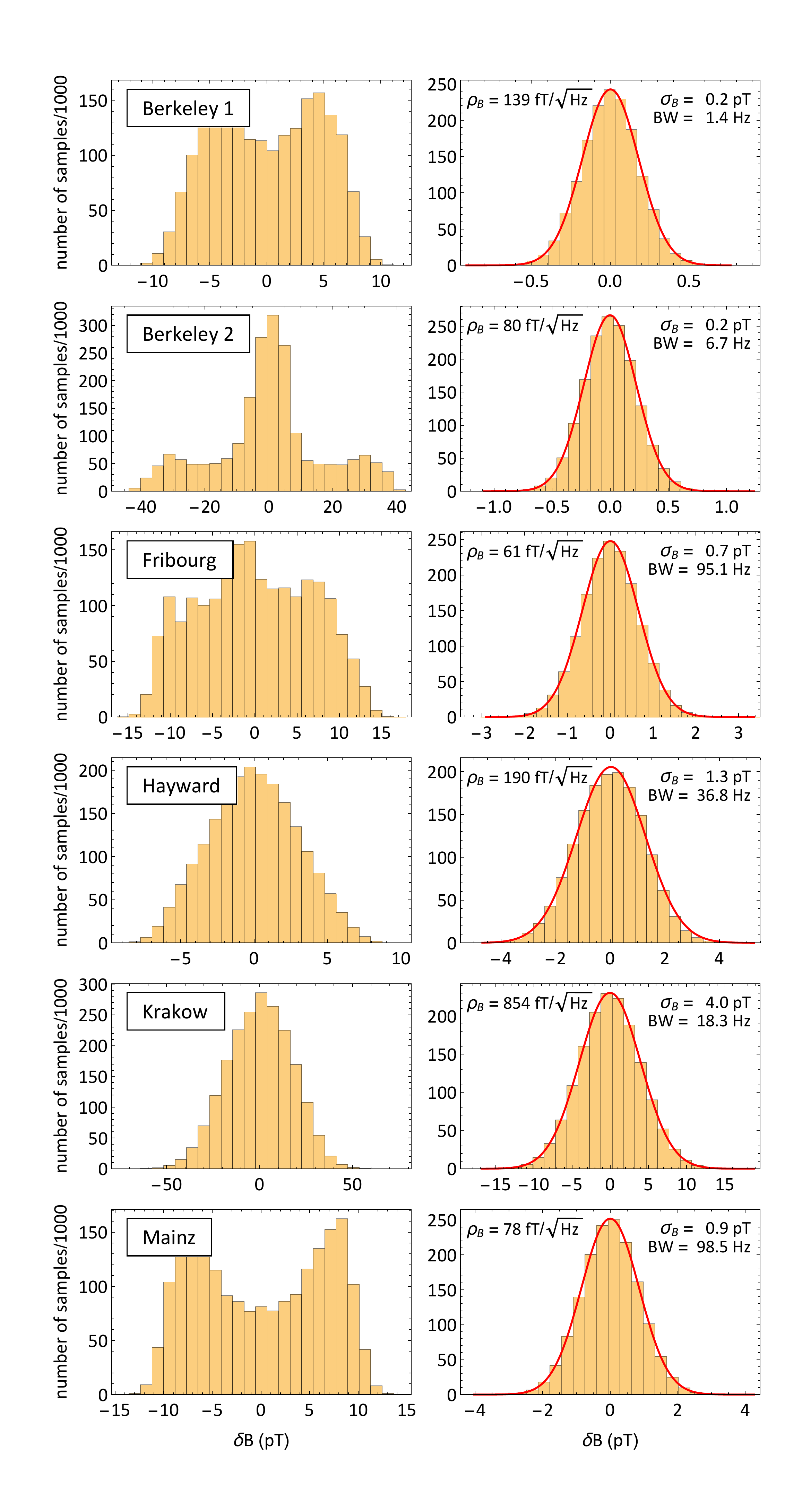}
\caption{Left: Histograms of the discrete field readings in the raw time series. Right: Histograms of filtered data and fitted Gaussian distributions (see text for details).
}
\label{Fig:Histograms1h}
\end{figure*}
Many foreseen approaches to analysis of GNOME data across the network to search for correlated transient signals are based on the assumption of a Gaussian distribution of the sampled field readings.
For this reason, we present in Fig.~\ref{Fig:Histograms1h} histograms of the signal amplitudes for the same data set used in Secs.~\ref{sec:psd} and \ref{sec:asd}.
The left column shows the histograms of the raw data (time series of Fig.~\ref{Fig:FFTs}).
Hayward and Krakow feature a bell-shaped distribution, the Berkeley 1 and Mainz data have a shape that is typical for harmonic oscillations, while Berkeley 2 and Fribourg show both features.
The right column of Fig.~\ref{Fig:Histograms1h} shows the histograms after filtering according to the procedure described and used in Sec.~\ref{sec:psd}.
The red curves superposed on those histograms represent fitted Gaussian distributions yielding the standard deviations denoted by $\sigma_B$ in the graphs, indicating that the filtered data are consistent with the assumption of Gaussian-distributed data.
When taking the magnetometer bandwidths into account, one can infer amplitude spectral densities $\rho_B\equiv\sigma_B/\sqrt{f_c}$ which are denoted in Fig.~\ref{Fig:Histograms1h} as well as shown as horizontal lines in the $\rho_B(f)$ plots of Fig.~\ref{Fig:FFTs}, which are consistent with the rms-average of the ($\approx$white) noise.

\section{Long-term data characteristics}
\label{Sec:long-term}

\begin{figure*}
\includegraphics[width=6.5 in]{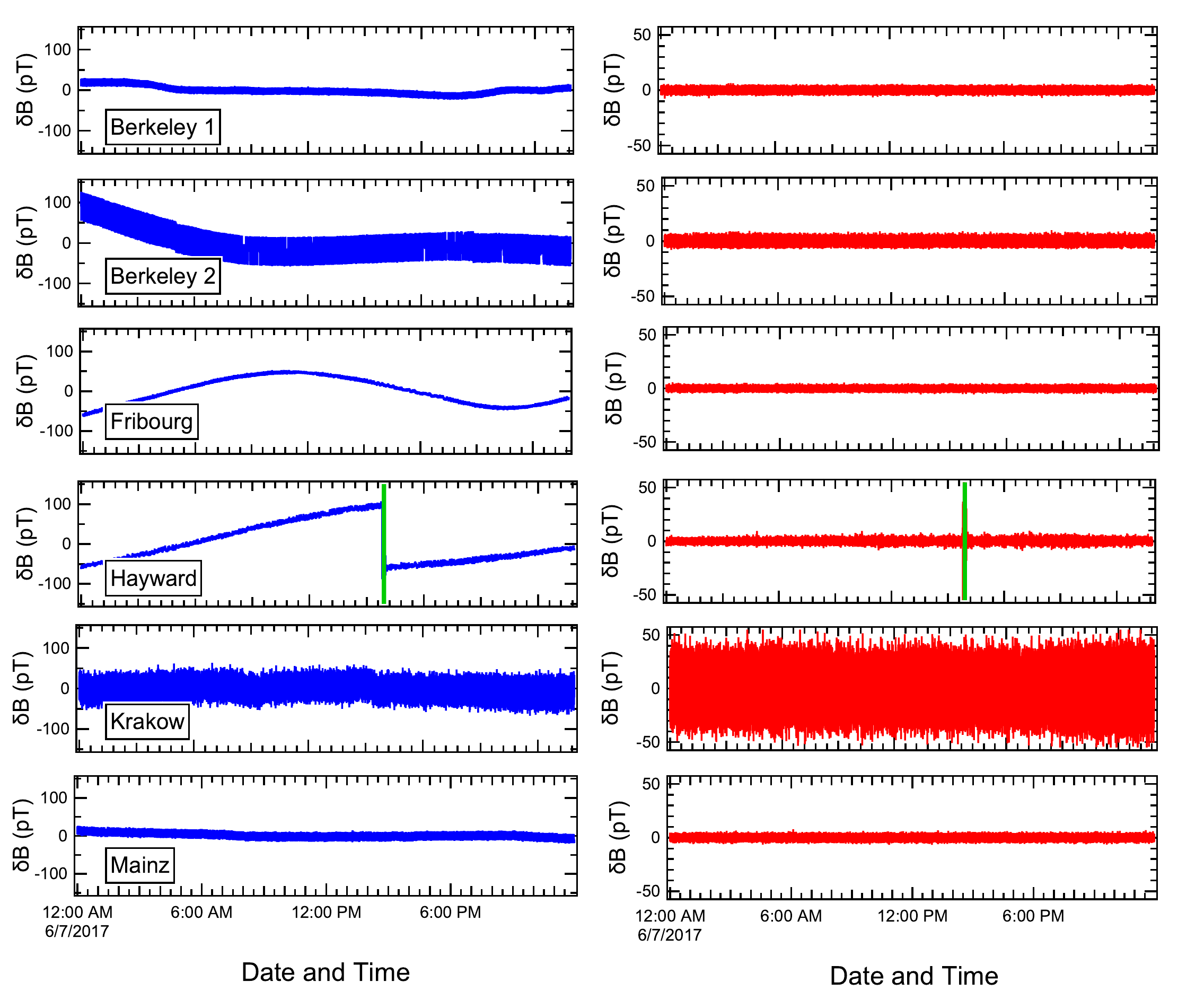}
\caption{One-day time series for all six magnetometers (7 June 2017).  The data are downsampled with one out of every 100 data points being displayed.  The plots on the left-hand side show the unfiltered raw magnetic-field data.  The plots on the right-hand side show the digitally filtered magnetic-field measurements.  All stations are digitally filtered with a high-pass filter as described in the text. Additional digital notch filters (described in text) are used to remove relatively large line oscillations for the filtered data on the right. The relatively large magnetic field jump observed by the Hayward station at around 2:00 pm UTC is highlighted in green to note that it was flagged as ``insane'' by the sanity monitor.}
\label{Fig:downsampledTimeSeries}
\end{figure*}

\begin{figure*}
\includegraphics[width=4.5 in]{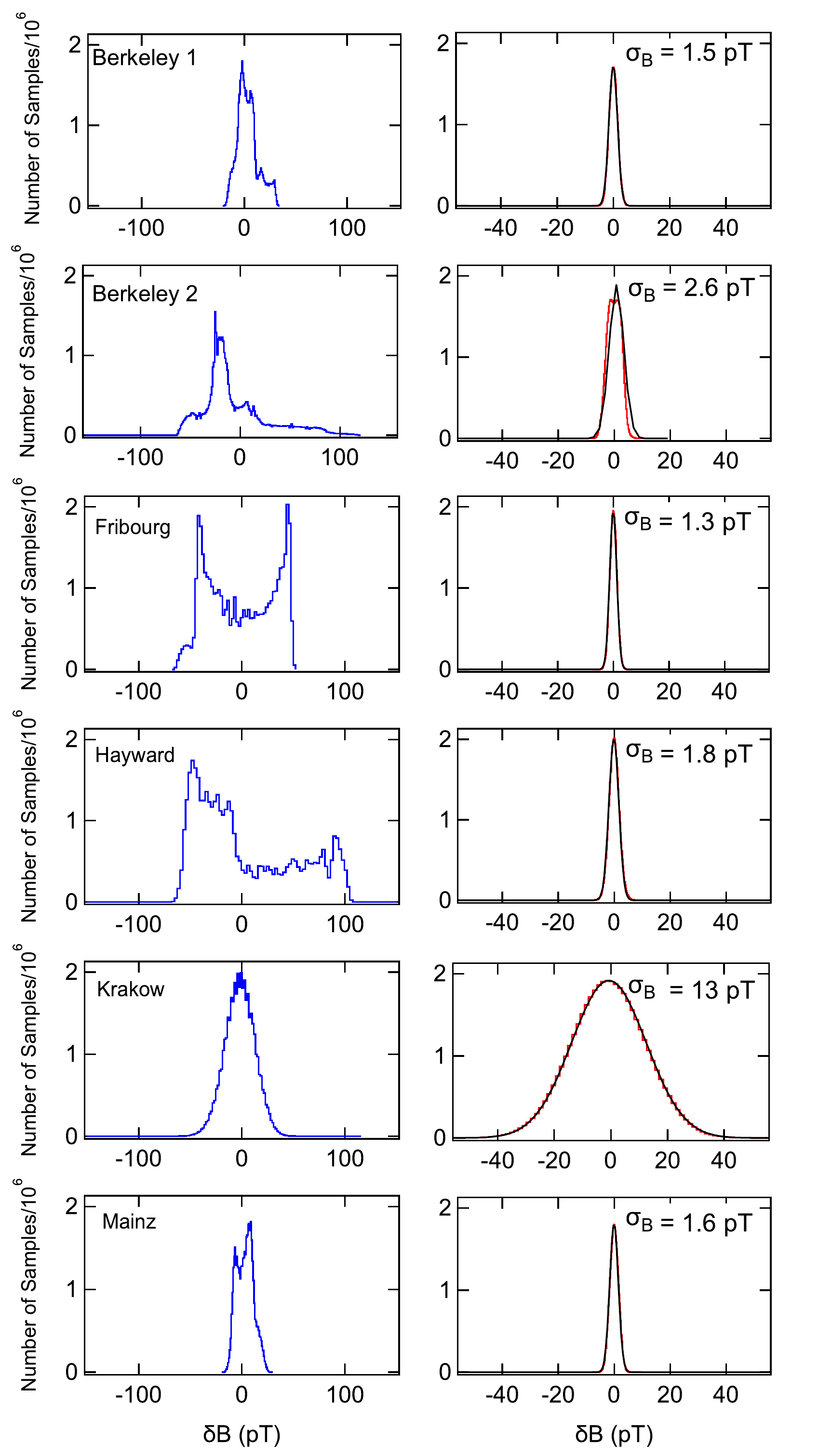}
\caption{Histograms of the magnetic field measurements over a one-day time period (7 June 2017) for all six GNOME magnetometers; these are the same data as are plotted in Fig.~\ref{Fig:downsampledTimeSeries}.  The plots on the left (blue) are the unfiltered data (centered about zero magnetic field for clarity). The plots on the right (red) are the digitally filtered data.  All stations are filtered with a high-pass filter and notch filters as described in the text. The black traces on the right-hand plots show Gaussian fits for the data.}
\label{Fig:Histograms}
\end{figure*}

Longer term characteristics of typical data for a 24-hour period from UTC 00:00:00 to 23:59:59 on 7 June 2017 are shown in Figs.~\ref{Fig:downsampledTimeSeries} and \ref{Fig:Histograms}.  As for all the data discussed in this work, the data were acquired at a rate of 500~S/s over this period.  A down-sampled display of these data are shown in Fig.~\ref{Fig:downsampledTimeSeries}, where only one of every 100 points is displayed for clarity of the plot.  As can be seen from the time series, all of the magnetometers have some level of drift over the course of a day. Also notable is the easily observable spike in the Hayward data occurring at roughly 2:00~pm UTC in Fig.~\ref{Fig:downsampledTimeSeries}. No other similar spikes are observed in the data from other stations anywhere within the roughly 40~s window where correlated transient signals from compact dark matter objects would be expected. Thus it can be concluded that the Hayward event must be attributable to some local phenomenon, illustrating the value of multiple, geographically separated detectors for suppression of false positives.

In fact, the apparent Hayward magnetic field jump at 2:00~pm on 7 June 2017 was flagged as ``insane'' by the Hayward sanity channel (Sec.~\ref{Sec:Sanity}). Subsequent investigation of the data logged by the sanity monitor indicate that this jump was caused by the probe laser losing and regaining lock during routine maintenance of the station. This incident thus also illustrates the utility of the sanity monitor for vetoing false positive signals.

As noted in Sec.~\ref{sec:psd}, to reduce the effects of drifts and other noise sources, digital filtering, for example, can be used in post-processing of GNOME data. Figure~\ref{Fig:downsampledTimeSeries} also shows data in the plots on the right-hand side that were digitally filtered by applying a high-pass single-pole Butterworth filter with a corner frequency of $10$ mHz in both the increasing and decreasing time sequence.  By applying the filter in this way, the resulting filter has linear phase.  In addition, three of the stations exhibit strong coherent oscillations;  Berkeley 1 station at 72.5 Hz, Berkeley 2 station at 60 Hz and 180 Hz, and the Mainz station at 50 Hz. For these stations six-order Butterworth filters were applied in both the increasing and decreasing time directions to eliminate these coherent oscillations. These filters had a 1-Hz wide stop band.

The effect of the filtering is clear in the time series: the long term drift of the measured magnetic field observed in the raw data by all stations is eliminated by the filters.  In addition, the noise level is reduced for the stations where notch filters were applied.

As discussed above, digital filtering allows for a clearer study of the overall short-term noise character of the signals as well.  Most of the data analysis techniques to be applied in the search for correlated transient signals incorporate an underlying assumption of Gaussian-distributed noise. Histograms of the data from each of the stations are shown in Fig.~\ref{Fig:Histograms}.  The filtering results in near Gaussian noise at long time scales for each of the stations.  Digital filtering in post-processing of GNOME data thus appears to be a suitable method to prepare data for further detailed analysis to search for correlated transient events. We note that the coherent oscillations of the Berkeley 1, Berkeley 2, and Mainz stations manifest as non-Gaussian distributions in the histograms for the unfiltered signals, with Berkeley 1 and Mainz having distributions matching what one would expect for coherent oscillations.  The histogram of Berkeley 2 shows a more complicated distribution that is the result of having strong oscillations at both 60 Hz and 180 Hz.

\section{Conclusion}

We have described the experimental setup for the Global Network of Optical Magnetometers to search for Exotic physics (GNOME), including the magnetometer setups, general characteristics related to the sensitivity to exotic fields, the GPS-disciplined data acquisition system, the ``sanity'' monitor for identifying and flagging transient signals due to magnetometer component failure and/or environmental perturbations, as well as the data format, transfer, and storage infrastructure of the GNOME. The GNOME is designed to search for transient or otherwise time-dependent signals of astrophysical origin heralding exotic physics, in particular exotic fields that couple to atomic spins.  The sensitivities and noise characteristics of the magnetometers were studied and discussed. This characterization will inform future efforts to analyze data collected during Science Runs searching for various types of exotic physics, for example, terrestrial encounters with compact dark matter objects \cite{Pos13,Kim18AxionStars}.

\appendix
\section{Experimental setups of individual GNOME magnetometers}
\label{Appendix:ExptSetups}

\subsection{Berkeley station 1}
\label{Sec:Berkeley01Setup}

\begin{figure*}
\includegraphics[width=.9\textwidth]{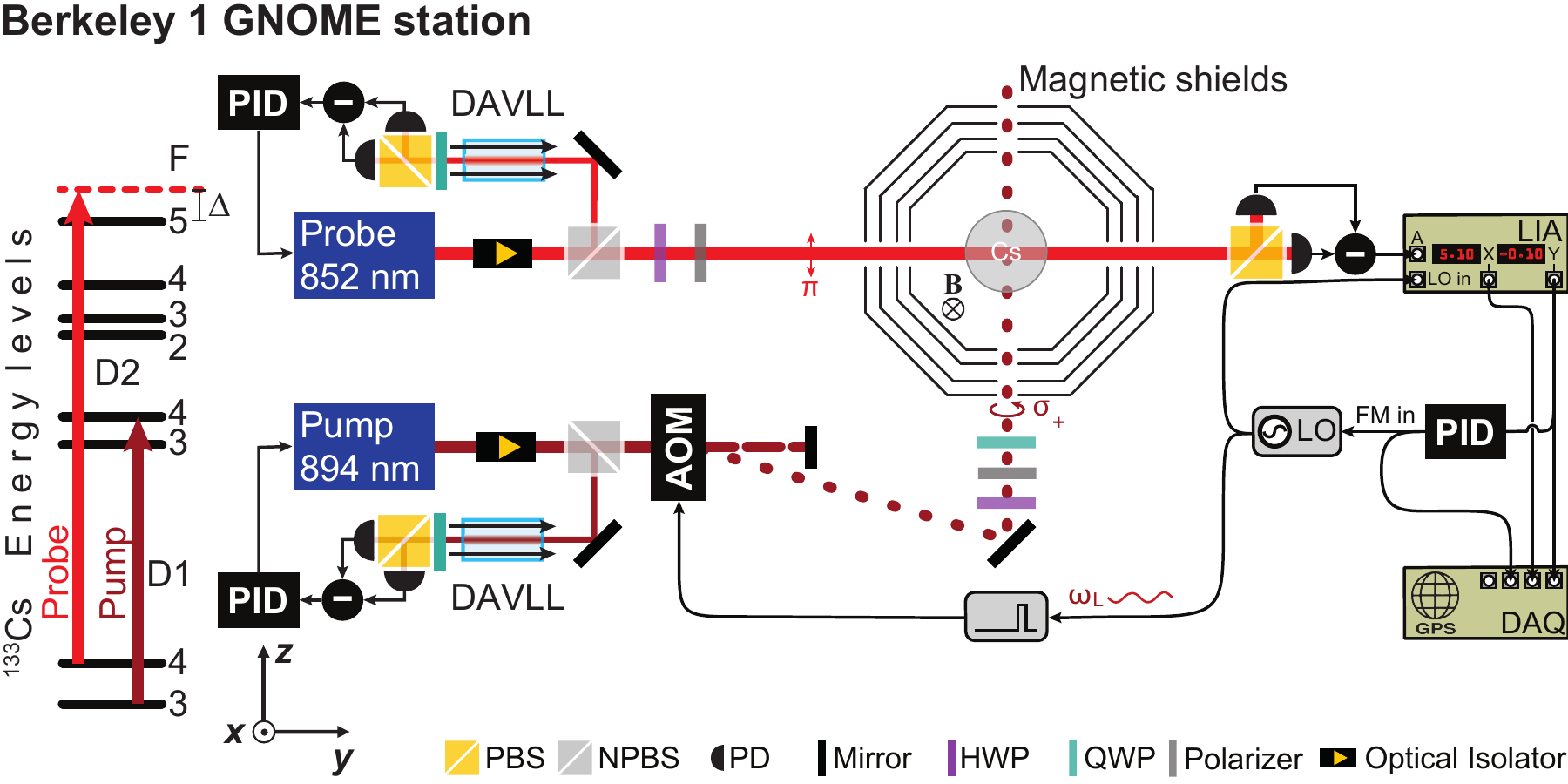}
\caption{Schematic diagram of the experimental setup for the Berkeley 1 GNOME magnetometer. The $\hat{\mb{y}}$ axis is along the probe beam and $\hat{\mb{z}}$ points up along the vertical direction. Notation is the same as in Fig.~\ref{Fig:ExptSetupFribourg}, with the following new terms: DAVLL = dichroic atomic vapor laser lock system \cite{yashchuk2000davll} and AOM = acousto-optic modulator. See Refs.~\cite{2013:budker,budker2002resonant} for discussion of the magnetic shield system design.}
\label{fig:Berkeley1SetupExample}
\end{figure*}

The Berkeley 1 GNOME magnetometer, shown schematically in Fig.~\ref{fig:Berkeley1SetupExample}, is located in a second-floor laboratory on the University of California at Berkeley campus. It is a two-beam amplitude-modulated nonlinear magneto-optical rotation (AM NMOR) magnetometer, similar in design to the Bell-Bloom scheme (see Refs.~\cite{2013:budker,budker2002resonant} for reviews). At the center of the apparatus is a cylindrical antirelaxation-coated Cs vapor cell (length $\approx 5$~cm, diameter $\approx 5$~cm). The Cs atoms contained in the cell have a spin-relaxation time of $\approx 0.7$~s, which enables generation of narrow magneto-optical resonances facilitating high-sensitivity magnetometry. The cell is contained within a custom four-layer mu-metal magnetic shield. The layers are separated by foam insulation to also improve thermal stability. The cell and shields are at ambient room temperature (typically $\sim 20 - 22^\circ$C). A set of internal magnetic-field coils (not shown) allows control of uniform field components along three orthogonal directions as well as all first-order gradients. The currents to the coils are generated by a custom supply which can provide up to 150 mA (Magnicon GmbH). The current supply is housed in a temperature-stabilized enclosure and exhibits a relative drift of $\sim 10^{-7}$ over 100 seconds. As noted in Table~\ref{Table:BasicCharacteristics}, a leading magnetic field $\mb{B}$ of magnitude $489~{\rm nT}$ is applied to the atoms along the horizontal $-\hat{\mb{x}}$ direction (orthogonal to both the pump- and probe-laser beam paths), corresponding to a Larmor frequency $\omega_L/(2\pi) \approx 1710~{\rm Hz}$.

The Cs atoms are synchronously optically pumped using circularly polarized light resonant with the 894~nm Cs D1 $F=3 \rightarrow F'=4$ transition (time-averaged power $\approx 17$~\textmu\textrm{W}). The pump beam both creates atomic spin polarization (orientation) in the $F=4$ hyperfine level and optically pumps atoms from the $F=3$ hyperfine level to the $F=4$ hyperfine level to increase the signal. The pump beam is generated with a distributed feedback (DFB) laser whose wavelength is locked to the center of the Doppler-broadened atomic resonance using a dichroic atomic vapor laser lock (DAVLL) system. The pump beam is amplitude-modulated at $\omega\ts{mod}/(2\pi) \approx 1710~{\rm Hz}$ by using an acousto-optic modulator (AOM). The AOM is driven with a local oscillator (LO) tuned with PID control electronics to the Larmor frequency $\omega_L$ based on the measured signal from the lock-in amplifier (LIA) monitoring the probe signal.

The linearly-polarized probe beam is tuned to the high-frequency wing of the 852~nm Cs D2 line, nearest to the $F=4 \rightarrow F'=5$ transition (power $\approx 10$~\textmu\textrm{W}). The optical frequency is stabilized with a second DAVLL system; the detuning and intensity of the probe beam are optimized for the largest signal with minimal power broadening. After exiting the vapor cell and magnetic shield assembly, the probe beam passes through a polarizing beam splitting cube (PBS) and the resulting beams are directed into an balanced photoreceiver. The output of the photoreceiver is sent to the lock-in amplifier (LIA, Stanford Research Systems SR830). The demodulated signal from the LIA is used both to keep the LO tuned to the magnetic resonance frequency as well as a measure for the magnetic field. Presently, the Berkeley 1 magnetometer is optimized for sensitivity rather than bandwidth, and hence the bandwidth of the magnetometer as shown in Fig.~\ref{fig:bandwidth} roughly corresponds to the transverse spin-relaxation time in the Cs vapor. In future runs, the phase-locked loop that keeps the AOM frequency tuned to $\omega_L$ will be optimized for a $\sim$100~Hz bandwidth.

\subsection{Berkeley station 2}
\label{Sec:Berkeley02Setup}

The Berkeley station 2 is located in a lab in the same building and on the same floor as Berkeley 1 but it is set up for an orthogonal sensitive axis. At the core of the Berkeley 2 magnetometer is a cylindrical, antirelaxation-coated Cs vapor cell enclosed within five layers of custom mu-metal magnetic shielding. The spaces between the magnetic shielding layers are filled with foam insulation to improve thermal stability. All the measurements are performed at ambient temperature.

Multiple coils are mounted inside the innermost layer of the magnetic shield system allowing magnetic fields and gradients to be applied to the cell. This setup only utilizes the $\hat{\mb{x}}$ direction coil, orthogonal to the probe beam propagation direction. The field is produced by a DC current source (Krohn-Hite Model 523) providing a current of 20.29 mA (corresponding to $\approx $6.7~kHz Larmor frequency).

The pump beam is generated by a DFB laser and tuned to the Cs D1 $F=3 \rightarrow F'=4$ transition and has a time-averaged power of 50 \textmu W as measured before the entrance of the shield. A laser diode driver (Wavelength Electronics LFI-4502) is used to drive the current of the pump diode laser, and a temperature controller (Thorlabs TED 200) to control the temperature of the diode. The beam emerging from the pump laser is split into two different paths: one path for the feedback loop that stabilizes the diode laser frequency to the Cs D1 $F=3 \rightarrow F'=4$ transition, the other path for pumping the Cs atoms in the antirelaxation-coated cell at the center of the magnetic-shield system. The pump beam is circularly polarized along the $-\hat{\mb{y}}$ direction.

\begin{figure*}
\includegraphics[width=.9\textwidth]{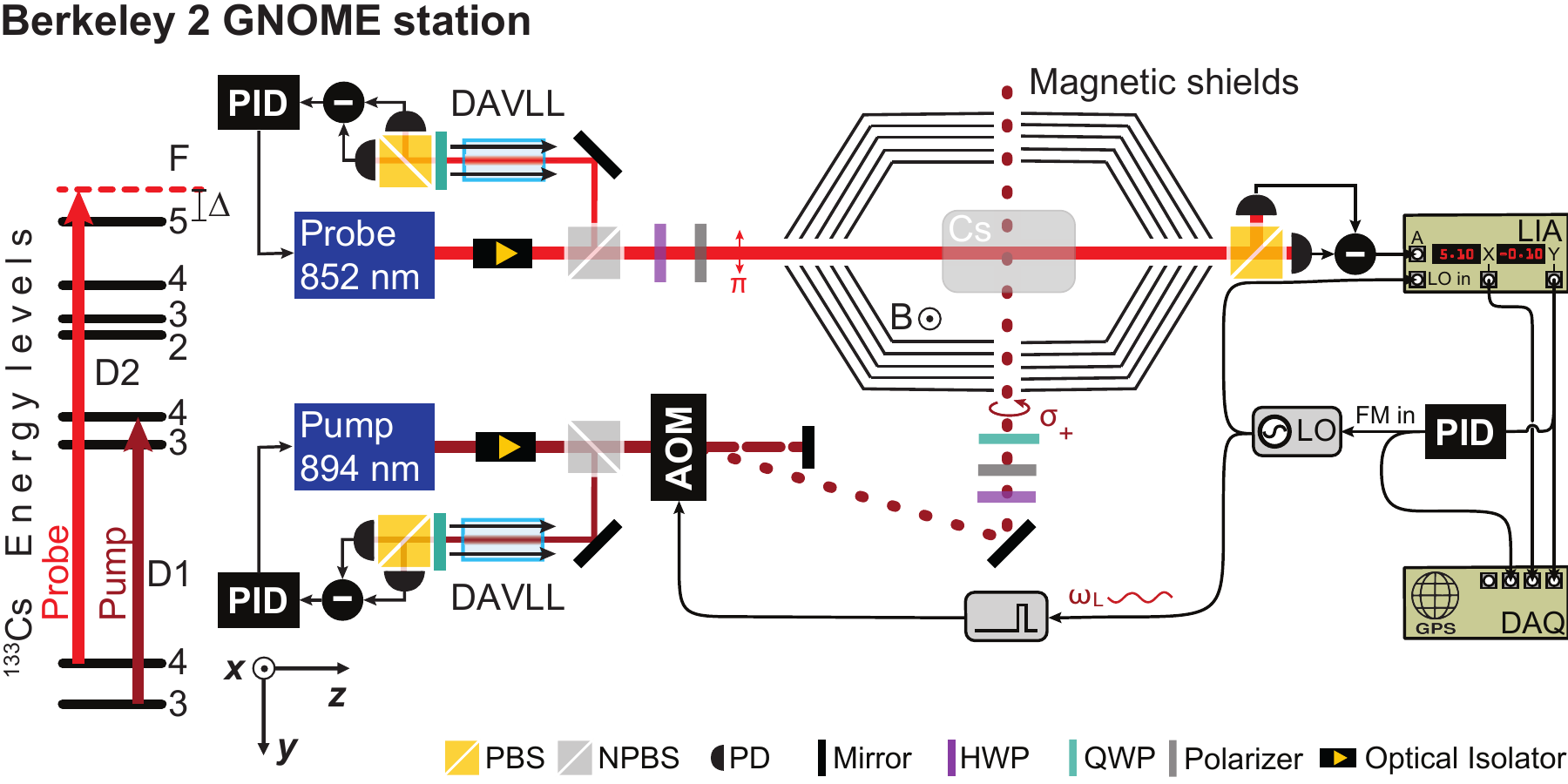}
\caption{Schematic diagram of the experimental setup for the Berkeley 2 GNOME magnetometer. The $\hat{\mb{z}}$ axis is along the cylindrical axis of the shield and $\hat{\mb{x}}$ points up along the vertical direction.Notation is the same as in Figs.~\ref{Fig:ExptSetupFribourg} and \ref{fig:Berkeley1SetupExample}. See Refs.~\cite{2013:budker,budker2002resonant} for discussion of the magnetic shield system design.}
\label{fig:Berkeley2Setup}
\end{figure*}

The differential signal coming out of the DAVLL polarimeter board is connected to the PID controller (Stanford Research Systems SIM960), from which the output signal is sent back to the laser diode driver (Wavelength Electronics, model LFI 4500). The pump beam is amplitude modulated with duty cycle of about 20\%.

A laser diode combi-controller (Thorlabs ITC 502) is used to drive the current and control the temperature of the probe diode laser. The probe laser is tuned to the Cs D2 $F=4 \rightarrow F'=5$ transition. The probe beam has a time-averaged power of 30~\textmu W at the entrance the shield and is linearly polarized, propagating along the $z$ axis as shown in Fig.~\ref{fig:Berkeley2Setup}. A DAVLL-based feedback loop stabilizes the probe laser frequency. Optical probing of the spin-precession and the phase-locked loop feedback control of the pump-modulation frequency are carried out using the same techniques as employed in the Berkeley 1 GNOME magnetometer setup.

\subsection{Hayward station}
\label{Sec:HaywardSetup}

\begin{figure*}
\includegraphics[width=0.9\textwidth]{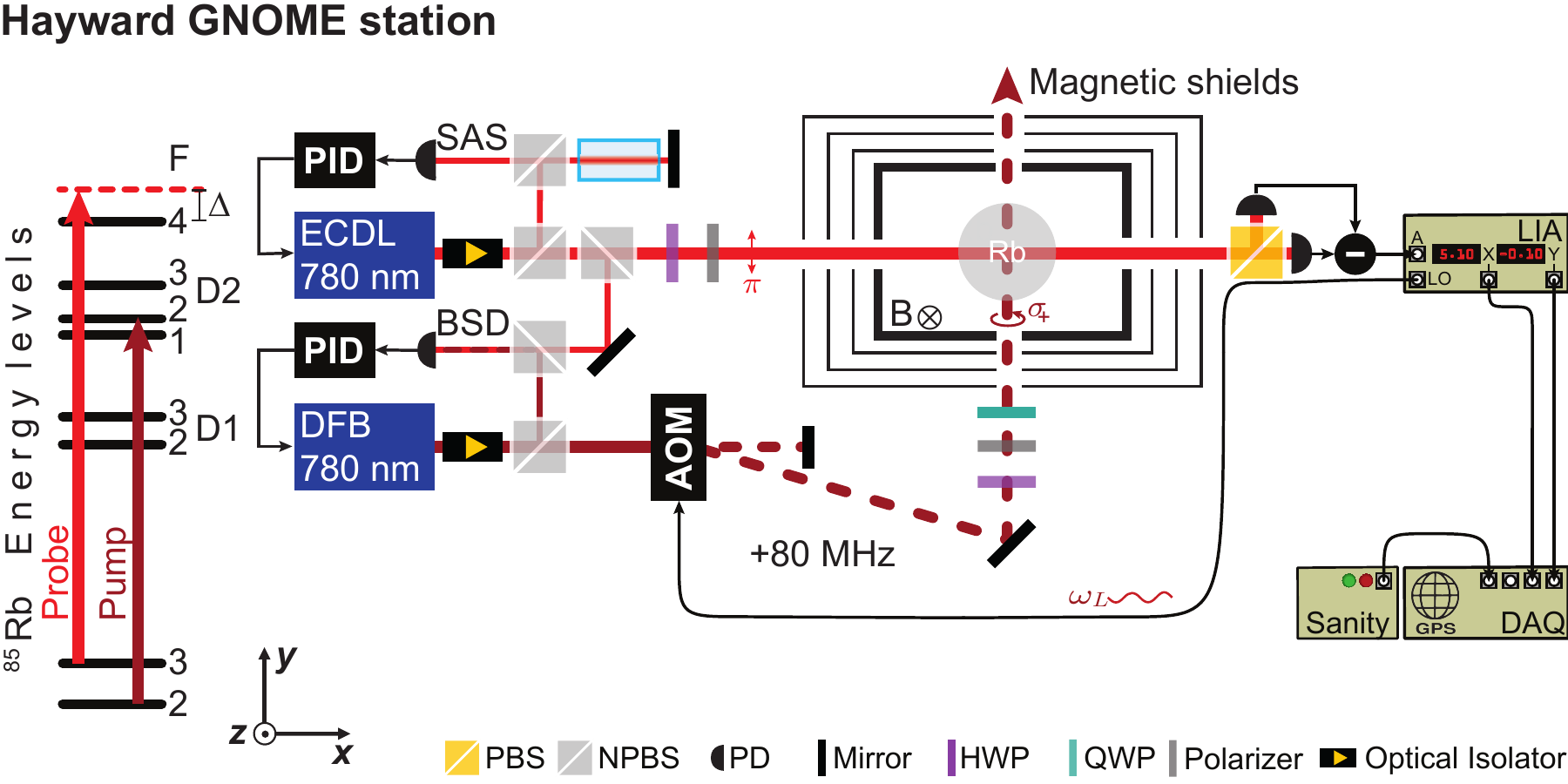}
\caption{Schematic diagram of the experimental setup for the Hayward GNOME magnetometer. Notation is the same as in Figs.~\ref{Fig:ExptSetupFribourg}~and \ref{fig:Berkeley1SetupExample}, with the following new terms: DFB = distributed feedback laser system and BSD =  beat signal detector (and associated electronics). The leading magnetic field is oriented down with respect to the vertical direction in the laboratory ($-\hat{\mb{z}}$).}
\label{Fig:ExptSetupHayward}
\end{figure*}

The Hayward GNOME station magnetometer (Fig.~\ref{Fig:ExptSetupHayward}, located in a ground-floor laboratory at California State University - East Bay), is a two-beam amplitude-modulated nonlinear magneto-optical rotation (AM NMOR) magnetometer using a natural isotopic mixture of Rb atoms ($\approx 72\%$ $^{85}$Rb, $\approx 28\%$ $^{87}$Rb) contained within a spherical paraffin-coated vapor cell ($\approx 5~{\rm cm}$ in diameter) at ambient room temperature ($\approx 24^\circ{\rm C}$). The total Rb vapor density within the paraffin-coated cell is $\approx 7 \times 10^9~{\rm atoms/cm^3}$. The cell is contained within a four-layer cylindrical magnetic shield manufactured by TwinLeaf LLC (TwinLeaf MS-1F), consisting of three outer layers made of mu-metal and an inner layer made of ferrite. A set of internal magnetic field coils allows control of uniform field components along three orthogonal directions ($\hat{\mb{x}}$, $\hat{\mb{y}}$, and $\hat{\mb{z}}$, where $\hat{\mb{y}}$ is along the cylindrical axis of the shield and $\hat{\mb{z}}$ points up along the vertical direction) as well as all first-order gradients.

The pump beam, which propagates along $\hat{\mb{y}}$, is generated by a distributed feedback (DFB) laser system (Toptica DFB pro). The probe beam, which propagates in the $\hat{\mb{x}}$ direction, is generated by a single external-cavity diode laser (ECDL) system (New Focus Vortex TLM 7000). Both pump and probe beams are resonant with the 780~nm Rb D2 transition. The linearly-polarized probe beam is locked to the high-frequency wing of the $^{85}$Rb $F=3 \rightarrow F'$ transition ($\approx 200~{\rm MHz}$ from the Doppler-broadened line center) by using a saturated-absorption spectroscopy (SAS) setup. The circularly-polarized pump beam is locked to the center of the $^{85}$Rb $F=2 \rightarrow F'$ transition by detecting the beat note between the pump and probe beams and referencing the beat note to a voltage-controlled oscillator (VCO). The pump beam is amplitude-modulated at $\Omega\ts{mod}/(2\pi) \approx 6975~{\rm Hz}$ by using an acousto-optic modulator (Intra-Action ATM-801A1 driven at 80 MHz by an Intra-Action Model DFE frequency synthesizer); the sinusoidal modulation is produced by the internal oscillator of the lock-in amplifier (Signal Recovery model 7265).  The pump beam power (unmodulated) is $\approx 400$ \textmu W and the probe beam power is $\approx 200$ \textmu W, and both beams have a diameter of $\approx 1~{\rm mm}$. After exiting the vapor cell and magnetic shield assembly, the probe beam passes through a Wollaston polarizing beam splitting cube and the resulting beams are directed into an autobalanced photoreceiver (New Focus Nirvana Model 2007). The output of the photoreceiver is then sent to the lock-in amplifier set to a time constant of 640 \textmu s with a filter roll-off of -6~dB/octave. The entire apparatus is on an optical table with passively air-damped supports whose bases are in buckets of sand.

A leading magnetic field applied via the coils is directed along the vertical direction ($-\hat{\mb{z}}$) and tuned so that the Larmor frequency $\omega_L$ matches the modulation frequency. The magnetic field is passively monitored by observing the out-of-phase output of the lock-in amplifier, which has a dispersive dependence on the magnetic field along $z$. Both the in-phase and out-of-phase lock-in signals are sent to the GPS DAQ box (Sec.~\ref{Sec:GPS-DAQ}); after passing through a low-pass filter with a 1~kHz pass-band (Thorlabs EF110: 1~kHz is the 3~dB point, there is 40~dB suppression above 3~kHz) in order to avoid aliasing of higher-frequency noise. The output of the sanity monitor (Sec.~\ref{Sec:Sanity}) is also sent to the GPS DAQ. The sanity system monitors the laser lock signals as well as the in-phase output of the LIA. The magnetometer has first-order sensitivity along the vertical direction but only second-order sensitivity for directions transverse to $z$. Data from the GPS DAQ box is continuously transferred to a PC and uploaded to the central server in Mainz, Germany.

\subsection{Krakow station}
\label{Sec:KrakowSetup}

\begin{figure*}
\includegraphics[width=.9\textwidth]{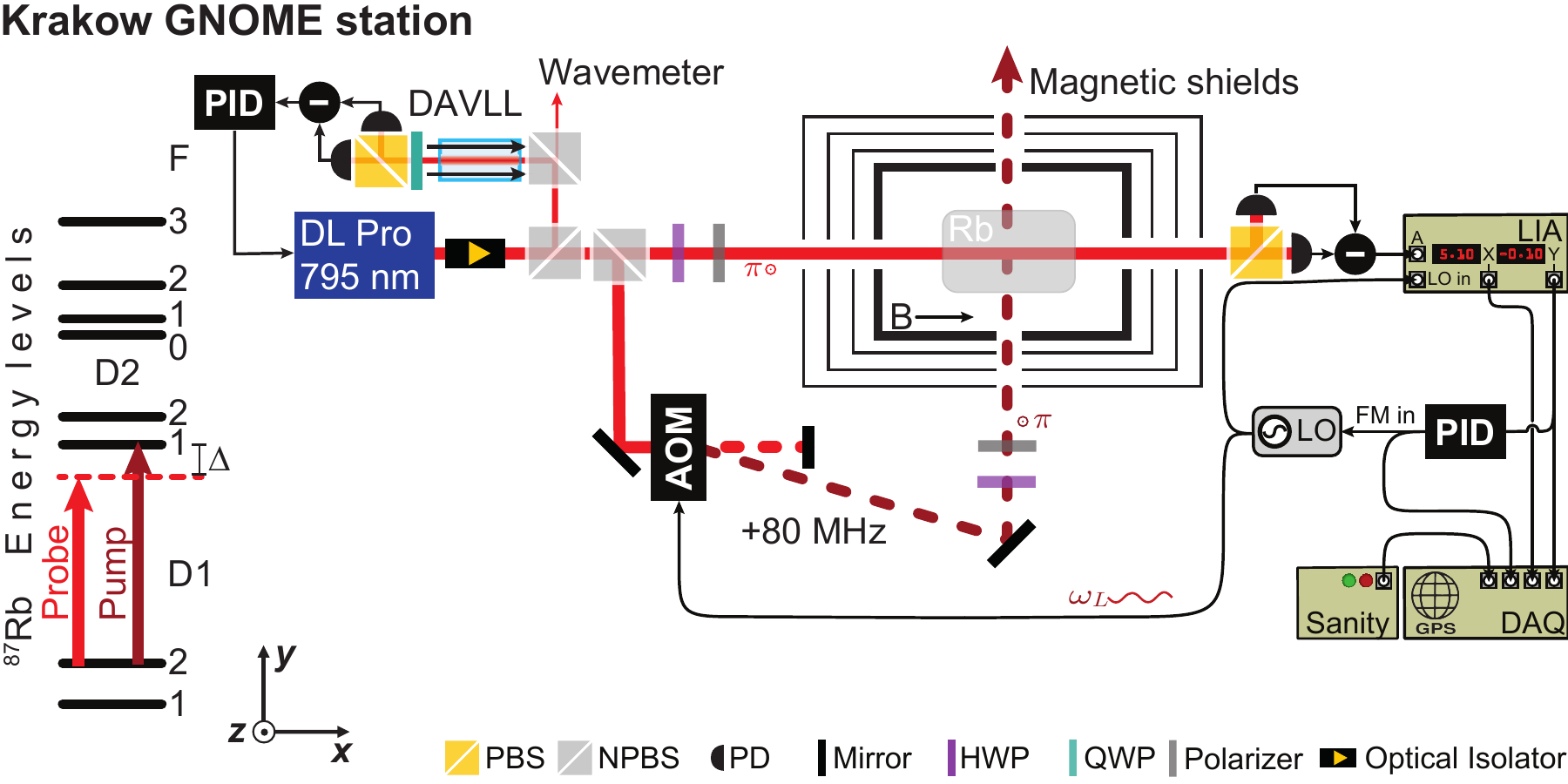}
\caption{Schematic diagram of the experimental setup for the Krakow GNOME magnetometer. Notation is the same as in Figs.~\ref{Fig:ExptSetupFribourg}~and \ref{fig:Berkeley1SetupExample}.}
\label{fig:KrakowSetup}
\end{figure*}

The Krakow GNOME station, located at the first floor of the Faculty of Physics, Astronomy, and Applied Computer Science of the Jagiellonian University, operates using a two-beam AM NMOR magnetometer. A schematic diagram of the magnetometer is shown in Fig.~\ref{fig:KrakowSetup}. The heart of the magnetometer is an isotopically enriched sample of $^{87}$Rb contained in a paraffin-coated spherical PYREX glass cell of 2~cm in diameter with a lockable stem that houses a rubidium droplet. The cell is operated at room temperature ($\approx 25^\circ$C), roughly corresponding to a concentration of $5\times 10^9$~atoms/cm$^3$. The cell is placed inside a cylindrical magnetic shield (Twinleaf MS-1F) consisting of three layers of mu-metal and an innermost ferrite layer. A set of magnetic-field coils installed inside the shield are used to generate a magnetic field homogeneous in all three directions as well as to minimize all first-order magnetic field gradients. The coils allow compensation of residual fields (up to linear gradients) and generation of a leading magnetic field along $\hat{\mb{x}}$.

The cell is illuminated with light emitted from the extended cavity diode laser (Toptica DL pro). The light is tuned to the low-frequency wing of the Doppler-broadened $F=2\rightarrow F'=1$ transition of the Rb D1 line (795~nm). Its frequency is stabilized with a dichroic atomic vapor laser lock (DAVLL), exploiting a buffer-gas-filled micro-cell \cite{Pustelny2016Dichroic}, enabling a broad locking range. The light frequency is monitored with a SAS system and wavemeter. The light is split into two beams of roughly similar intensities. One of the beams (the pump) passes through an acousto-optical modulator (AOM), which operates in the first-order diffraction regime. The amplitude of the AOM’s 80-MHz acoustic wave is sinusoidally modulated with 100\% modulation depth. This results in modulation of intensity of light directed into the first order. Interaction of light with the acoustic wave in the AOM also leads to a shift of light frequency by 80~MHz. Thereby the pump is tuned closer to the center of the Rb $F=2\rightarrow F'=1$ transition. After adjustment of its intensity (with half-wave plate and crystal polarizer) $z$-polarized light propagates through the vapor along $\hat{\mb{y}}$. The other beam (the probe) is also polarized along $z$ and, after adjusting its intensity, traverses the vapor cell along $\hat{\mb{x}}$.  After the medium, the polarization state of the probe is determined using a home-made balanced polarimeter, consisting of a Wollaston prism (WP) and two photodiodes with associated electronics. The photodiode difference signal provides information about the probe-light-polarization rotation angle.

Inside the shield, the rubidium vapor is subjected to an $\hat{\mb{x}}$-oriented magnetic field of $\approx 1.1$~\textmu T. Since the field corresponds to the Larmor frequency of 8.1~kHz, to fulfill the AM NMOR resonance condition ($\omega_m=2\omega_L$), the pump is modulated at 16.2~kHz. This enables generation of a macroscopic dynamic spin polarization of the atomic medium, which modulates the probe light polarization.

The polarimeter difference signal is demodulated at the first harmonic of the modulation frequency $\omega_m$ with a LIA (Stanford Research SR830). The signal is measured with a time constant of 1~ms and filtered with -12~dB/octave filter. In this arrangement the in-phase component of the signal is fed to a PID system, which controls a generator used for intensity modulation. Such an approach allows operation of the magnetometer in the phase-stabilizing mode, which accounts for slow and fast field drifts by modifying the modulation frequency. A PID output signal is also fed to the GPS DAQ system (DMTechnologies GDL100), where it is stored (Sec.~\ref{Sec:GPS-DAQ}). Precalibration of the signal enables one to convert the PID control voltage into magnetic field. Additionally, the quadrature signal is fed into the sanity monitor (Sec.~\ref{Sec:Sanity}). The limits set on the box provide that the signal does not deteriorate below some level (below 80\% of maximum signal). Additionally, the sanity system monitors the DAVLL signal to ensure that the laser does not lose its lock.

\subsection{Mainz station}
\label{Sec:MainzSetup}

The Mainz GNOME station is a two-beam AM NMOR magnetometer. The sensor is based on an evacuated paraffin-coated $^{87}$Rb vapor cell placed in a magnetically shielded environment. This system is located in the basement of the Helmholtz Institute Mainz in a temperature-stabilized room.

\begin{figure*}[!ht]
 \centering
\includegraphics[width=\textwidth]{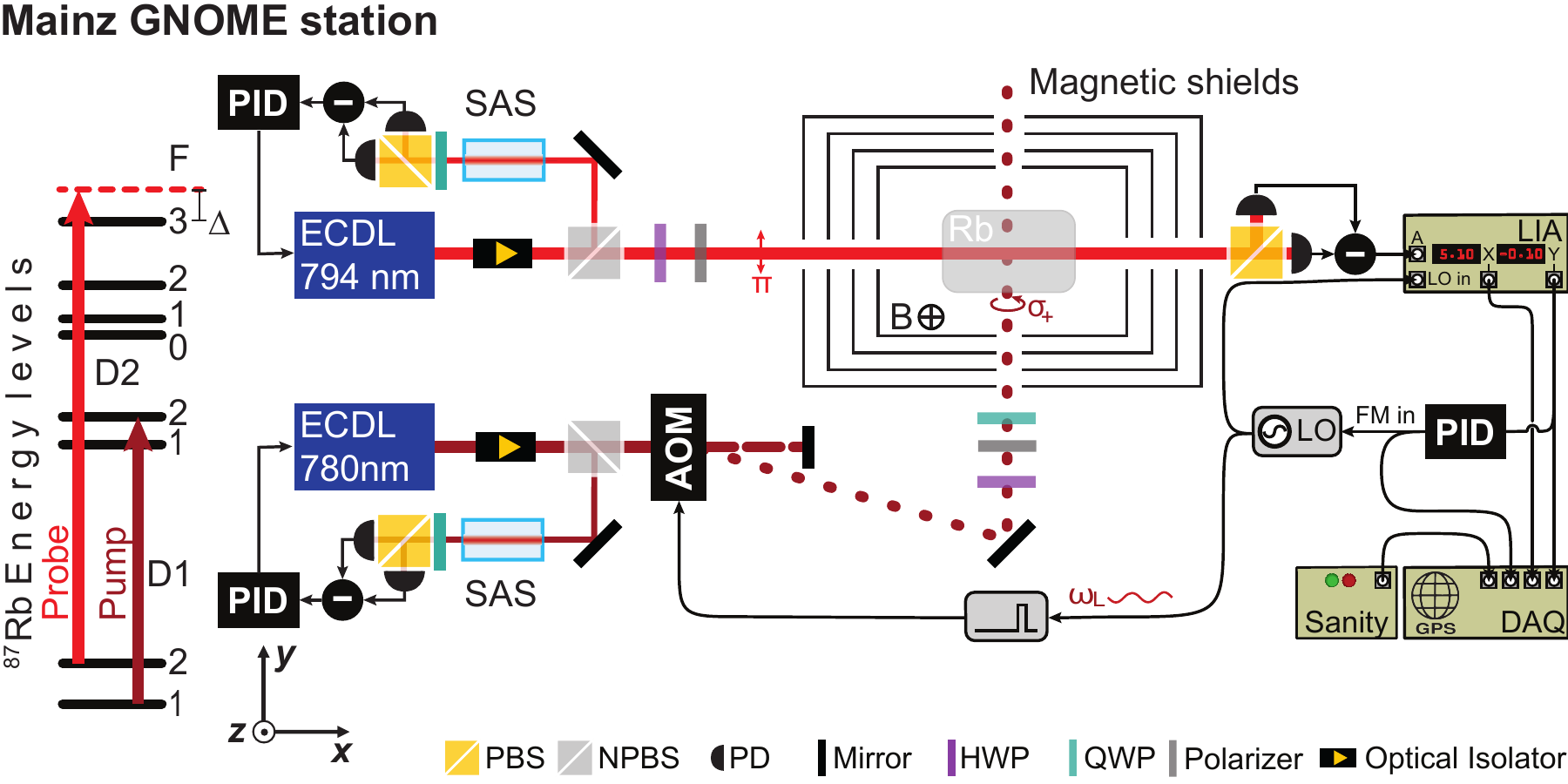}
\caption{Schematic diagram of the experimental setup for the Mainz GNOME magnetometer. Notation is the same as in Figs.~\ref{Fig:ExptSetupFribourg}~and \ref{fig:Berkeley1SetupExample}.}
\label{Diagram}
\end{figure*}

The paraffin-coated $^{87}$Rb vapor cell at the center of the apparatus has a cylindrical shape with length of 5~cm and diameter of 5~cm. The measurements are performed at a stabilized room temperature of 21$^{\circ}$C. At this temperature the atomic density is 8.2\,$\times$\,10$^9$\,atoms/cm$^3$. The cell is placed inside a custom four-layer mu-metal shield. A set of three square magnetic coils is located inside the shields with coil axes aligned so that they are mutually perpendicular. The coil that defines the $\hat{\mb{z}}$-axis, oriented perpendicular to the ground, also establishes the direction of the leading magnetic field which has a magnitude of 525\,pT pointing opposite to $\hat{\mb{z}}$. The current source for the coils is the model SEL-1 manufactured by Magnicon.

The pump beam is produced by an external-cavity diode laser (ECDL) manufactured by Vitawave. Its optical frequency is locked to resonance with the $^{87}$Rb $5~^2\textnormal{S}_{1/2}, \, F=1 \rightarrow 5~^2\textnormal{P}_{1/2}, \, F'=2$ transition using a custom-made SAS setup. In order to produce an error signal, the light beam is frequency-modulated at 10$\,$kHz. The beam propagates in the $\hat{\mb{y}}$-direction and before interacting with the cell volume, the light is circularly polarized. To excite the spin precession, the beam is periodically pulsed with a pulse duration of 1\,\textmu s from 0 to $4.1$\,\ mW using an AOM manufactured by Isomet. The AOM is driven by a function generator (Tektronix AFG2021) which allows modulation of the pulsing frequency $\Omega_{\text{mod}}$ according to a voltage input to a VCO. This feature is used to lock the pulsing frequency to the magnetic resonance frequency using a PID controller based on the out-of-phase output of the lock-in amplifier. This feedback loop keeps the pulsing frequency tuned to the magnetic resonance frequency ($\Omega_{\text{mod}}=\Omega_L=2 \pi \times 3791 \,$Hz, where $\Omega_L$ is the Larmor frequency). In this configuration, the measurement of the magnetic field is given by the output voltage of the PID controller (proportional to frequency changes of the local oscillator).

The linearly polarized probe beam, which propagates in the $\hat{\mb{x}}$-direction, is generated by a ECDL system (Moglabs CEL002). The optical frequency is locked to the D2 crossover between the $5~^2\textnormal{S}_{1/2},~F=3 \rightarrow 5~^2\textnormal{P}_{3/2},~F'=3 ~\textnormal{and}~ F'=4$ transitions of $^{85}$Rb using the SAS system; the error signal is produced by frequency modulation at 100~kHz. This locking point is detuned by +1.2\,GHz from the $^{87}$Rb $5~^2\textnormal{S}_{1/2}, \, F=2 \rightarrow 5~^2\textnormal{P}_{1/2}, \, F'=2$ transition. The probe beam measures the population dynamics in the Zeeman sublevels of the $5~^2\textnormal{S}_{1/2}, \, F=2$ state. The beam power is fixed to 368\,~\textmu W. The optical rotation produced by the precessing spins is measured with a differential polarimeter composed of a Wollaston prism and a balanced photoreceiver (Thorlabs PDB210A).

The sinusoidal signal at $\Omega_L$ measured with the balanced photoreceiver is processed with a two-phase LIA (Standford Research Systems SR830). In order to cut before the Nyquist frequency of 250 Hz, the bandwidth is limited to 150\,Hz for magnetic-field detection using the LIA. It is set to a time constant of 300~\textmu s with roll-off -24~dB/octave. The reference signal is given by the local oscillator that produces the pump-pulse frequency. The out-of-phase output is used to lock the local oscillator frequency on resonance while the in-phase component confirms that the resonance condition is matched.

In order to monitor the magnetometer operational status, the amplitude of the absorption peak and the error signal from the SAS as well as the magnetic resonance amplitude are fed to the sanity monitor (Sec.~\ref{Sec:Sanity}). The output of the sanity monitor is sent to the GPS DAQ box (Sec.~\ref{Sec:GPS-DAQ}).

Magnetic-field changes are measured through the PID output voltage that controls the local oscillator. This signal scales with the magnetic field as 714.3\,pT/V. In order to avoid aliasing due to the limited sampling rate of 500\,samples/s the signal is further filtered before recording it with the GPS DAQ. The filter used is a second-order Butterworth filter with a cut-off frequency of 200\,Hz.

\acknowledgments

The authors are sincerely grateful to Andrei Derevianko, Przemys\l{}aw W\l{}odarczyk, Dongok Kim, Yun Shin, Ibrahim Sulai, and Joshua Smith for enlightening discussions. This work was supported by the U.S. National Science Foundation under grants PHY-1707875 and PHY-1707803, grant No. 200021 172686 of the Swiss National Science Foundation, and the European Research Council under the European Union's Horizon 2020 Research and Innovative Program under Grant agreement No. 695405. H.G. and X.P. acknowledge support from National Key R\&D Program of China (QIT), National Hi-Tech R\&D Program (863) of China (QIT) and the National Natural Science Foundation of China (61531003 and 61571018). W.L. acknowledges support from the China Scholarship Council (CSC). The author list is arranged alphabetically.

\newpage

\bibliography{references}

\begin{thebibliography}{54}
\expandafter\ifx\csname natexlab\endcsname\relax\def\natexlab#1{#1}\fi
\expandafter\ifx\csname bibnamefont\endcsname\relax
  \def\bibnamefont#1{#1}\fi
\expandafter\ifx\csname bibfnamefont\endcsname\relax
  \def\bibfnamefont#1{#1}\fi
\expandafter\ifx\csname citenamefont\endcsname\relax
  \def\citenamefont#1{#1}\fi
\expandafter\ifx\csname url\endcsname\relax
  \def\url#1{\texttt{#1}}\fi
\expandafter\ifx\csname urlprefix\endcsname\relax\def\urlprefix{URL }\fi
\providecommand{\bibinfo}[2]{#2}
\providecommand{\eprint}[2][]{\url{#2}}

\bibitem[{\citenamefont{Budker and Jackson~Kimball}(2013)}]{2013:budker}
\bibinfo{author}{\bibfnamefont{D.}~\bibnamefont{Budker}} \bibnamefont{and}
  \bibinfo{author}{\bibfnamefont{D.~F.} \bibnamefont{Jackson~Kimball}},
  \emph{\bibinfo{title}{Optical Magnetometry}} (\bibinfo{publisher}{Cambridge
  University Press}, \bibinfo{year}{2013}).

\bibitem[{\citenamefont{Safronova et~al.}(2018)\citenamefont{Safronova, Budker,
  DeMille, Jackson~Kimball, Derevianko, and Clark}}]{Saf18RMP}
\bibinfo{author}{\bibfnamefont{M.~S.} \bibnamefont{Safronova}},
  \bibinfo{author}{\bibfnamefont{D.}~\bibnamefont{Budker}},
  \bibinfo{author}{\bibfnamefont{D.}~\bibnamefont{DeMille}},
  \bibinfo{author}{\bibfnamefont{D.~F.} \bibnamefont{Jackson~Kimball}},
  \bibinfo{author}{\bibfnamefont{A.}~\bibnamefont{Derevianko}},
  \bibnamefont{and} \bibinfo{author}{\bibfnamefont{C.~W.} \bibnamefont{Clark}},
  \bibinfo{journal}{Rev. Mod. Phys.} \textbf{\bibinfo{volume}{90}},
  \bibinfo{pages}{025008} (\bibinfo{year}{2018}).

\bibitem[{\citenamefont{Venema et~al.}(1992)\citenamefont{Venema, Majumder,
  Lamoreaux, Heckel, and Fortson}}]{1992:venema}
\bibinfo{author}{\bibfnamefont{B.~J.} \bibnamefont{Venema}},
  \bibinfo{author}{\bibfnamefont{P.~K.} \bibnamefont{Majumder}},
  \bibinfo{author}{\bibfnamefont{S.~K.} \bibnamefont{Lamoreaux}},
  \bibinfo{author}{\bibfnamefont{B.~R.} \bibnamefont{Heckel}},
  \bibnamefont{and} \bibinfo{author}{\bibfnamefont{E.~N.}
  \bibnamefont{Fortson}}, \bibinfo{journal}{Phys. Rev. Lett.}
  \textbf{\bibinfo{volume}{68}}, \bibinfo{pages}{135} (\bibinfo{year}{1992}).

\bibitem[{\citenamefont{Heckel et~al.}(2008)\citenamefont{Heckel, Adelberger,
  Cramer, Cook, Schlamminger, and Schmidt}}]{2008:heckel}
\bibinfo{author}{\bibfnamefont{B.~R.} \bibnamefont{Heckel}},
  \bibinfo{author}{\bibfnamefont{E.~G.} \bibnamefont{Adelberger}},
  \bibinfo{author}{\bibfnamefont{C.~E.} \bibnamefont{Cramer}},
  \bibinfo{author}{\bibfnamefont{T.~S.} \bibnamefont{Cook}},
  \bibinfo{author}{\bibfnamefont{S.}~\bibnamefont{Schlamminger}},
  \bibnamefont{and} \bibinfo{author}{\bibfnamefont{U.}~\bibnamefont{Schmidt}},
  \bibinfo{journal}{Phys. Rev. D} \textbf{\bibinfo{volume}{78}},
  \bibinfo{pages}{092006} (\bibinfo{year}{2008}).

\bibitem[{\citenamefont{Jackson~Kimball
  et~al.}(2017)\citenamefont{Jackson~Kimball, Dudley, Li, Patel, and
  Valdez}}]{2017:kimball}
\bibinfo{author}{\bibfnamefont{D.~F.} \bibnamefont{Jackson~Kimball}},
  \bibinfo{author}{\bibfnamefont{J.}~\bibnamefont{Dudley}},
  \bibinfo{author}{\bibfnamefont{Y.}~\bibnamefont{Li}},
  \bibinfo{author}{\bibfnamefont{D.}~\bibnamefont{Patel}}, \bibnamefont{and}
  \bibinfo{author}{\bibfnamefont{J.}~\bibnamefont{Valdez}},
  \bibinfo{journal}{Phys. Rev. D} \textbf{\bibinfo{volume}{96}},
  \bibinfo{pages}{075004} (\bibinfo{year}{2017}).

\bibitem[{\citenamefont{Hunter et~al.}(2013)\citenamefont{Hunter, Gordon, Peck,
  Ang, and Lin}}]{Hun13}
\bibinfo{author}{\bibfnamefont{L.}~\bibnamefont{Hunter}},
  \bibinfo{author}{\bibfnamefont{J.}~\bibnamefont{Gordon}},
  \bibinfo{author}{\bibfnamefont{S.}~\bibnamefont{Peck}},
  \bibinfo{author}{\bibfnamefont{D.}~\bibnamefont{Ang}}, \bibnamefont{and}
  \bibinfo{author}{\bibfnamefont{J.-F.} \bibnamefont{Lin}},
  \bibinfo{journal}{Science} \textbf{\bibinfo{volume}{339}},
  \bibinfo{pages}{928} (\bibinfo{year}{2013}).

\bibitem[{\citenamefont{Youdin et~al.}(1996)\citenamefont{Youdin, Krause,
  Jagannathan, Hunter, and Lamoreaux}}]{You96}
\bibinfo{author}{\bibfnamefont{A.~N.} \bibnamefont{Youdin}},
  \bibinfo{author}{\bibfnamefont{D.}~\bibnamefont{Krause}},
  \bibinfo{author}{\bibfnamefont{K.}~\bibnamefont{Jagannathan}},
  \bibinfo{author}{\bibfnamefont{L.~R.} \bibnamefont{Hunter}},
  \bibnamefont{and} \bibinfo{author}{\bibfnamefont{S.~K.}
  \bibnamefont{Lamoreaux}}, \bibinfo{journal}{Phys. Rev. Lett.}
  \textbf{\bibinfo{volume}{77}}, \bibinfo{pages}{2170} (\bibinfo{year}{1996}).

\bibitem[{\citenamefont{Chu et~al.}(2013)\citenamefont{Chu, Dennis, Fu, Gao,
  Khatiwada, Laskaris, Li, Smith, Snow, Yan et~al.}}]{Chu13}
\bibinfo{author}{\bibfnamefont{P.-H.} \bibnamefont{Chu}},
  \bibinfo{author}{\bibfnamefont{A.}~\bibnamefont{Dennis}},
  \bibinfo{author}{\bibfnamefont{C.~B.} \bibnamefont{Fu}},
  \bibinfo{author}{\bibfnamefont{H.}~\bibnamefont{Gao}},
  \bibinfo{author}{\bibfnamefont{R.}~\bibnamefont{Khatiwada}},
  \bibinfo{author}{\bibfnamefont{G.}~\bibnamefont{Laskaris}},
  \bibinfo{author}{\bibfnamefont{K.}~\bibnamefont{Li}},
  \bibinfo{author}{\bibfnamefont{E.}~\bibnamefont{Smith}},
  \bibinfo{author}{\bibfnamefont{W.~M.} \bibnamefont{Snow}},
  \bibinfo{author}{\bibfnamefont{H.}~\bibnamefont{Yan}}, \bibnamefont{et~al.},
  \bibinfo{journal}{Phys. Rev. D} \textbf{\bibinfo{volume}{87}},
  \bibinfo{pages}{011105(R)} (\bibinfo{year}{2013}).

\bibitem[{\citenamefont{Vasilakis et~al.}(2009)\citenamefont{Vasilakis, Brown,
  Kornack, and Romalis}}]{Vas09}
\bibinfo{author}{\bibfnamefont{G.}~\bibnamefont{Vasilakis}},
  \bibinfo{author}{\bibfnamefont{J.~M.} \bibnamefont{Brown}},
  \bibinfo{author}{\bibfnamefont{T.~W.} \bibnamefont{Kornack}},
  \bibnamefont{and} \bibinfo{author}{\bibfnamefont{M.~V.}
  \bibnamefont{Romalis}}, \bibinfo{journal}{Phys. Rev. Lett.}
  \textbf{\bibinfo{volume}{103}}, \bibinfo{pages}{261801}
  (\bibinfo{year}{2009}).

\bibitem[{\citenamefont{Lee et~al.}(2018)\citenamefont{Lee, Almasi, and
  Romalis}}]{Lee18}
\bibinfo{author}{\bibfnamefont{J.}~\bibnamefont{Lee}},
  \bibinfo{author}{\bibfnamefont{A.}~\bibnamefont{Almasi}}, \bibnamefont{and}
  \bibinfo{author}{\bibfnamefont{M.}~\bibnamefont{Romalis}},
  \bibinfo{journal}{Phys. Rev. Lett.} \textbf{\bibinfo{volume}{120}},
  \bibinfo{pages}{161801} (\bibinfo{year}{2018}).

\bibitem[{\citenamefont{Smiciklas et~al.}(2011)\citenamefont{Smiciklas, Brown,
  Cheuk, Smullin, and Romalis}}]{Smi11}
\bibinfo{author}{\bibfnamefont{M.}~\bibnamefont{Smiciklas}},
  \bibinfo{author}{\bibfnamefont{J.~M.} \bibnamefont{Brown}},
  \bibinfo{author}{\bibfnamefont{L.~W.} \bibnamefont{Cheuk}},
  \bibinfo{author}{\bibfnamefont{S.~J.} \bibnamefont{Smullin}},
  \bibnamefont{and} \bibinfo{author}{\bibfnamefont{M.~V.}
  \bibnamefont{Romalis}}, \bibinfo{journal}{Phys. Rev. Lett.}
  \textbf{\bibinfo{volume}{107}}, \bibinfo{pages}{171604}
  (\bibinfo{year}{2011}).

\bibitem[{\citenamefont{Gemmel et~al.}(2010)\citenamefont{Gemmel, Heil, Karpuk,
  Lenz, Sobolev, Tullney, Burghoff, Kilian, Knappe-Gr{\"u}neberg, M{\"u}ller
  et~al.}}]{Gem10}
\bibinfo{author}{\bibfnamefont{C.}~\bibnamefont{Gemmel}},
  \bibinfo{author}{\bibfnamefont{W.}~\bibnamefont{Heil}},
  \bibinfo{author}{\bibfnamefont{S.}~\bibnamefont{Karpuk}},
  \bibinfo{author}{\bibfnamefont{K.}~\bibnamefont{Lenz}},
  \bibinfo{author}{\bibfnamefont{Y.}~\bibnamefont{Sobolev}},
  \bibinfo{author}{\bibfnamefont{K.}~\bibnamefont{Tullney}},
  \bibinfo{author}{\bibfnamefont{M.}~\bibnamefont{Burghoff}},
  \bibinfo{author}{\bibfnamefont{W.}~\bibnamefont{Kilian}},
  \bibinfo{author}{\bibfnamefont{S.}~\bibnamefont{Knappe-Gr{\"u}neberg}},
  \bibinfo{author}{\bibfnamefont{W.}~\bibnamefont{M{\"u}ller}},
  \bibnamefont{et~al.}, \bibinfo{journal}{Phys. Rev. D}
  \textbf{\bibinfo{volume}{82}}, \bibinfo{pages}{111901}
  (\bibinfo{year}{2010}).

\bibitem[{\citenamefont{Pospelov et~al.}(2013)\citenamefont{Pospelov, Pustelny,
  Ledbetter, Jackson~Kimball, Gawlik, and Budker}}]{Pos13}
\bibinfo{author}{\bibfnamefont{M.}~\bibnamefont{Pospelov}},
  \bibinfo{author}{\bibfnamefont{S.}~\bibnamefont{Pustelny}},
  \bibinfo{author}{\bibfnamefont{M.~P.} \bibnamefont{Ledbetter}},
  \bibinfo{author}{\bibfnamefont{D.~F.} \bibnamefont{Jackson~Kimball}},
  \bibinfo{author}{\bibfnamefont{W.}~\bibnamefont{Gawlik}}, \bibnamefont{and}
  \bibinfo{author}{\bibfnamefont{D.}~\bibnamefont{Budker}},
  \bibinfo{journal}{Phys. Rev. Lett.} \textbf{\bibinfo{volume}{110}},
  \bibinfo{pages}{021803} (\bibinfo{year}{2013}).

\bibitem[{\citenamefont{Jackson~Kimball
  et~al.}(2018)\citenamefont{Jackson~Kimball, Budker, Eby, Pospelov, Pustelny,
  Scholtes, Stadnik, Weis, and Wickenbrock}}]{Kim18AxionStars}
\bibinfo{author}{\bibfnamefont{D.~F.} \bibnamefont{Jackson~Kimball}},
  \bibinfo{author}{\bibfnamefont{D.}~\bibnamefont{Budker}},
  \bibinfo{author}{\bibfnamefont{J.}~\bibnamefont{Eby}},
  \bibinfo{author}{\bibfnamefont{M.}~\bibnamefont{Pospelov}},
  \bibinfo{author}{\bibfnamefont{S.}~\bibnamefont{Pustelny}},
  \bibinfo{author}{\bibfnamefont{T.}~\bibnamefont{Scholtes}},
  \bibinfo{author}{\bibfnamefont{Y.~V.} \bibnamefont{Stadnik}},
  \bibinfo{author}{\bibfnamefont{A.}~\bibnamefont{Weis}}, \bibnamefont{and}
  \bibinfo{author}{\bibfnamefont{A.}~\bibnamefont{Wickenbrock}},
  \bibinfo{journal}{Phys. Rev. D} \textbf{\bibinfo{volume}{97}},
  \bibinfo{pages}{043002} (\bibinfo{year}{2018}).

\bibitem[{\citenamefont{Budker and Derevianko}(2015)}]{budker2015data}
\bibinfo{author}{\bibfnamefont{D.}~\bibnamefont{Budker}} \bibnamefont{and}
  \bibinfo{author}{\bibfnamefont{A.}~\bibnamefont{Derevianko}},
  \bibinfo{journal}{PHYSICS TODAY} \textbf{\bibinfo{volume}{68}},
  \bibinfo{pages}{9} (\bibinfo{year}{2015}).

\bibitem[{\citenamefont{Anderson et~al.}(2001)\citenamefont{Anderson, Brady,
  Creighton, and Flanagan}}]{And01}
\bibinfo{author}{\bibfnamefont{W.~G.} \bibnamefont{Anderson}},
  \bibinfo{author}{\bibfnamefont{P.~R.} \bibnamefont{Brady}},
  \bibinfo{author}{\bibfnamefont{J.~D.~E.} \bibnamefont{Creighton}},
  \bibnamefont{and} \bibinfo{author}{\bibfnamefont{E.~E.}
  \bibnamefont{Flanagan}}, \bibinfo{journal}{Phys. Rev. D}
  \textbf{\bibinfo{volume}{63}}, \bibinfo{pages}{042003}
  (\bibinfo{year}{2001}).

\bibitem[{\citenamefont{Allen et~al.}(2012)\citenamefont{Allen, Anderson,
  Brady, Brown, and Creighton}}]{All12}
\bibinfo{author}{\bibfnamefont{B.}~\bibnamefont{Allen}},
  \bibinfo{author}{\bibfnamefont{W.~G.} \bibnamefont{Anderson}},
  \bibinfo{author}{\bibfnamefont{P.~R.} \bibnamefont{Brady}},
  \bibinfo{author}{\bibfnamefont{D.~A.} \bibnamefont{Brown}}, \bibnamefont{and}
  \bibinfo{author}{\bibfnamefont{J.~D.~E.} \bibnamefont{Creighton}},
  \bibinfo{journal}{Phys. Rev. D} \textbf{\bibinfo{volume}{85}},
  \bibinfo{pages}{122006} (\bibinfo{year}{2012}).

\bibitem[{\citenamefont{Klimenko et~al.}(2016)\citenamefont{Klimenko, Vedovato,
  Drago, Salemi, Tiwari, Prodi, Lazzaro, Ackley, Tiwari, Da~Silva
  et~al.}}]{klimenko2016method}
\bibinfo{author}{\bibfnamefont{S.}~\bibnamefont{Klimenko}},
  \bibinfo{author}{\bibfnamefont{G.}~\bibnamefont{Vedovato}},
  \bibinfo{author}{\bibfnamefont{M.}~\bibnamefont{Drago}},
  \bibinfo{author}{\bibfnamefont{F.}~\bibnamefont{Salemi}},
  \bibinfo{author}{\bibfnamefont{V.}~\bibnamefont{Tiwari}},
  \bibinfo{author}{\bibfnamefont{G.}~\bibnamefont{Prodi}},
  \bibinfo{author}{\bibfnamefont{C.}~\bibnamefont{Lazzaro}},
  \bibinfo{author}{\bibfnamefont{K.}~\bibnamefont{Ackley}},
  \bibinfo{author}{\bibfnamefont{S.}~\bibnamefont{Tiwari}},
  \bibinfo{author}{\bibfnamefont{C.}~\bibnamefont{Da~Silva}},
  \bibnamefont{et~al.}, \bibinfo{journal}{Phys. Rev. D}
  \textbf{\bibinfo{volume}{93}}, \bibinfo{pages}{042004}
  (\bibinfo{year}{2016}).

\bibitem[{\citenamefont{Pustelny et~al.}(2013)\citenamefont{Pustelny,
  Jackson~Kimball, Pankow, Ledbetter, Wlodarczyk, Wcislo, Pospelov, Smith,
  Read, Gawlik et~al.}}]{2013:pustelny}
\bibinfo{author}{\bibfnamefont{S.}~\bibnamefont{Pustelny}},
  \bibinfo{author}{\bibfnamefont{D.~F.} \bibnamefont{Jackson~Kimball}},
  \bibinfo{author}{\bibfnamefont{C.}~\bibnamefont{Pankow}},
  \bibinfo{author}{\bibfnamefont{M.~P.} \bibnamefont{Ledbetter}},
  \bibinfo{author}{\bibfnamefont{P.}~\bibnamefont{Wlodarczyk}},
  \bibinfo{author}{\bibfnamefont{P.}~\bibnamefont{Wcislo}},
  \bibinfo{author}{\bibfnamefont{M.}~\bibnamefont{Pospelov}},
  \bibinfo{author}{\bibfnamefont{J.~R.} \bibnamefont{Smith}},
  \bibinfo{author}{\bibfnamefont{J.}~\bibnamefont{Read}},
  \bibinfo{author}{\bibfnamefont{W.}~\bibnamefont{Gawlik}},
  \bibnamefont{et~al.}, \bibinfo{journal}{Ann. Phys. (Berl.)}
  \textbf{\bibinfo{volume}{525}}, \bibinfo{pages}{659} (\bibinfo{year}{2013}).

\bibitem[{\citenamefont{Jackson~Kimball
  et~al.}(2016)\citenamefont{Jackson~Kimball, Dudley, Li, Thulasi, Pustelny,
  Budker, and Zolotorev}}]{Kim16}
\bibinfo{author}{\bibfnamefont{D.~F.} \bibnamefont{Jackson~Kimball}},
  \bibinfo{author}{\bibfnamefont{J.}~\bibnamefont{Dudley}},
  \bibinfo{author}{\bibfnamefont{Y.}~\bibnamefont{Li}},
  \bibinfo{author}{\bibfnamefont{S.}~\bibnamefont{Thulasi}},
  \bibinfo{author}{\bibfnamefont{S.}~\bibnamefont{Pustelny}},
  \bibinfo{author}{\bibfnamefont{D.}~\bibnamefont{Budker}}, \bibnamefont{and}
  \bibinfo{author}{\bibfnamefont{M.}~\bibnamefont{Zolotorev}},
  \bibinfo{journal}{Phys. Rev. D} \textbf{\bibinfo{volume}{94}},
  \bibinfo{pages}{082005} (\bibinfo{year}{2016}).

\bibitem[{\citenamefont{W{\l}odarczyk et~al.}(2014)\citenamefont{W{\l}odarczyk,
  Pustelny, Budker, and Lipi{\'n}ski}}]{Wlo14}
\bibinfo{author}{\bibfnamefont{P.}~\bibnamefont{W{\l}odarczyk}},
  \bibinfo{author}{\bibfnamefont{S.}~\bibnamefont{Pustelny}},
  \bibinfo{author}{\bibfnamefont{D.}~\bibnamefont{Budker}}, \bibnamefont{and}
  \bibinfo{author}{\bibfnamefont{M.}~\bibnamefont{Lipi{\'n}ski}},
  \bibinfo{journal}{Nucl. Instr. Meth. Phys. Res. A}
  \textbf{\bibinfo{volume}{763}}, \bibinfo{pages}{150} (\bibinfo{year}{2014}).

\bibitem[{\citenamefont{Roberts et~al.}(2017)\citenamefont{Roberts, Blewitt,
  Dailey, Murphy, Pospelov, Rollings, Sherman, Williams, and
  Derevianko}}]{Rob17}
\bibinfo{author}{\bibfnamefont{B.~M.} \bibnamefont{Roberts}},
  \bibinfo{author}{\bibfnamefont{G.}~\bibnamefont{Blewitt}},
  \bibinfo{author}{\bibfnamefont{C.}~\bibnamefont{Dailey}},
  \bibinfo{author}{\bibfnamefont{M.}~\bibnamefont{Murphy}},
  \bibinfo{author}{\bibfnamefont{M.}~\bibnamefont{Pospelov}},
  \bibinfo{author}{\bibfnamefont{A.}~\bibnamefont{Rollings}},
  \bibinfo{author}{\bibfnamefont{J.}~\bibnamefont{Sherman}},
  \bibinfo{author}{\bibfnamefont{W.}~\bibnamefont{Williams}}, \bibnamefont{and}
  \bibinfo{author}{\bibfnamefont{A.}~\bibnamefont{Derevianko}},
  \bibinfo{journal}{Nat. Commun.} \textbf{\bibinfo{volume}{8}},
  \bibinfo{pages}{1195} (\bibinfo{year}{2017}).

\bibitem[{\citenamefont{Carroll and Field}(1994)}]{Car94}
\bibinfo{author}{\bibfnamefont{S.~M.} \bibnamefont{Carroll}} \bibnamefont{and}
  \bibinfo{author}{\bibfnamefont{G.~B.} \bibnamefont{Field}},
  \bibinfo{journal}{Phys. Rev. D} \textbf{\bibinfo{volume}{50}},
  \bibinfo{pages}{3867} (\bibinfo{year}{1994}).

\bibitem[{\citenamefont{Arvanitaki et~al.}(2015)\citenamefont{Arvanitaki,
  Baryakhtar, and Huang}}]{Arv15}
\bibinfo{author}{\bibfnamefont{A.}~\bibnamefont{Arvanitaki}},
  \bibinfo{author}{\bibfnamefont{M.}~\bibnamefont{Baryakhtar}},
  \bibnamefont{and} \bibinfo{author}{\bibfnamefont{X.}~\bibnamefont{Huang}},
  \bibinfo{journal}{Phys. Rev. D} \textbf{\bibinfo{volume}{91}},
  \bibinfo{pages}{084011} (\bibinfo{year}{2015}).

\bibitem[{\citenamefont{Ellis et~al.}(2004)\citenamefont{Ellis, Mavromatos, and
  Westmuckett}}]{Ell04}
\bibinfo{author}{\bibfnamefont{J.}~\bibnamefont{Ellis}},
  \bibinfo{author}{\bibfnamefont{N.~E.} \bibnamefont{Mavromatos}},
  \bibnamefont{and}
  \bibinfo{author}{\bibfnamefont{M.}~\bibnamefont{Westmuckett}},
  \bibinfo{journal}{Phys. Rev. D} \textbf{\bibinfo{volume}{70}},
  \bibinfo{pages}{044036} (\bibinfo{year}{2004}).

\bibitem[{\citenamefont{Budker et~al.}(2014)\citenamefont{Budker, Graham,
  Ledbetter, Rajendran, and Sushkov}}]{Bud14}
\bibinfo{author}{\bibfnamefont{D.}~\bibnamefont{Budker}},
  \bibinfo{author}{\bibfnamefont{P.~W.} \bibnamefont{Graham}},
  \bibinfo{author}{\bibfnamefont{M.}~\bibnamefont{Ledbetter}},
  \bibinfo{author}{\bibfnamefont{S.}~\bibnamefont{Rajendran}},
  \bibnamefont{and} \bibinfo{author}{\bibfnamefont{A.~O.}
  \bibnamefont{Sushkov}}, \bibinfo{journal}{Phys. Rev. X}
  \textbf{\bibinfo{volume}{4}}, \bibinfo{pages}{021030} (\bibinfo{year}{2014}).

\bibitem[{\citenamefont{Derevianko and Pospelov}(2014)}]{Der14}
\bibinfo{author}{\bibfnamefont{A.}~\bibnamefont{Derevianko}} \bibnamefont{and}
  \bibinfo{author}{\bibfnamefont{M.}~\bibnamefont{Pospelov}},
  \bibinfo{journal}{Nat. Phys.} \textbf{\bibinfo{volume}{10}},
  \bibinfo{pages}{933} (\bibinfo{year}{2014}).

\bibitem[{\citenamefont{Wcis{\l}o et~al.}(2017)\citenamefont{Wcis{\l}o,
  Morzy{\'n}ski, Bober, Cygan, Lisak, Ciury{\l}o, and
  Zawada}}]{wcislo2017experimental}
\bibinfo{author}{\bibfnamefont{P.}~\bibnamefont{Wcis{\l}o}},
  \bibinfo{author}{\bibfnamefont{P.}~\bibnamefont{Morzy{\'n}ski}},
  \bibinfo{author}{\bibfnamefont{M.}~\bibnamefont{Bober}},
  \bibinfo{author}{\bibfnamefont{A.}~\bibnamefont{Cygan}},
  \bibinfo{author}{\bibfnamefont{D.}~\bibnamefont{Lisak}},
  \bibinfo{author}{\bibfnamefont{R.}~\bibnamefont{Ciury{\l}o}},
  \bibnamefont{and} \bibinfo{author}{\bibfnamefont{M.}~\bibnamefont{Zawada}},
  \bibinfo{journal}{Nat. Astron.} \textbf{\bibinfo{volume}{1}},
  \bibinfo{pages}{0009} (\bibinfo{year}{2017}).

\bibitem[{\citenamefont{Wcisło et~al.}(2018)}]{Wcislo2018ojh}
\bibinfo{author}{\bibfnamefont{P.}~\bibnamefont{Wcisło}} \bibnamefont{et~al.}
  (\bibinfo{year}{2018}), \eprint{1806.04762}.

\bibitem[{\citenamefont{Stadnik and Flambaum}(2015)}]{Sta15}
\bibinfo{author}{\bibfnamefont{Y.~V.} \bibnamefont{Stadnik}} \bibnamefont{and}
  \bibinfo{author}{\bibfnamefont{V.~V.} \bibnamefont{Flambaum}},
  \bibinfo{journal}{Phys. Rev. Lett.} \textbf{\bibinfo{volume}{114}},
  \bibinfo{pages}{161301} (\bibinfo{year}{2015}).

\bibitem[{\citenamefont{Stadnik and Flambaum}(2016)}]{Sta16}
\bibinfo{author}{\bibfnamefont{Y.~V.} \bibnamefont{Stadnik}} \bibnamefont{and}
  \bibinfo{author}{\bibfnamefont{V.~V.} \bibnamefont{Flambaum}},
  \bibinfo{journal}{Phys. Rev. A} \textbf{\bibinfo{volume}{93}},
  \bibinfo{pages}{063630} (\bibinfo{year}{2016}).

\bibitem[{\citenamefont{Jacobs et~al.}(2015)\citenamefont{Jacobs, Weltman, and
  Starkman}}]{Jac15}
\bibinfo{author}{\bibfnamefont{D.~M.} \bibnamefont{Jacobs}},
  \bibinfo{author}{\bibfnamefont{A.}~\bibnamefont{Weltman}}, \bibnamefont{and}
  \bibinfo{author}{\bibfnamefont{G.~D.} \bibnamefont{Starkman}},
  \bibinfo{journal}{Phys. Rev. D} \textbf{\bibinfo{volume}{91}},
  \bibinfo{pages}{115023} (\bibinfo{year}{2015}).

\bibitem[{\citenamefont{Arvanitaki et~al.}(2016)\citenamefont{Arvanitaki,
  Dimopoulos, and Van~Tilburg}}]{Arv16}
\bibinfo{author}{\bibfnamefont{A.}~\bibnamefont{Arvanitaki}},
  \bibinfo{author}{\bibfnamefont{S.}~\bibnamefont{Dimopoulos}},
  \bibnamefont{and}
  \bibinfo{author}{\bibfnamefont{K.}~\bibnamefont{Van~Tilburg}},
  \bibinfo{journal}{Phys. Rev. Lett.} \textbf{\bibinfo{volume}{116}},
  \bibinfo{pages}{031102} (\bibinfo{year}{2016}).

\bibitem[{\citenamefont{Branca et~al.}(2017)\citenamefont{Branca, Bonaldi,
  Cerdonio, Conti, Falferi, Marin, Mezzena, Ortolan, Prodi, Taffarello
  et~al.}}]{Bra17}
\bibinfo{author}{\bibfnamefont{A.}~\bibnamefont{Branca}},
  \bibinfo{author}{\bibfnamefont{M.}~\bibnamefont{Bonaldi}},
  \bibinfo{author}{\bibfnamefont{M.}~\bibnamefont{Cerdonio}},
  \bibinfo{author}{\bibfnamefont{L.}~\bibnamefont{Conti}},
  \bibinfo{author}{\bibfnamefont{P.}~\bibnamefont{Falferi}},
  \bibinfo{author}{\bibfnamefont{F.}~\bibnamefont{Marin}},
  \bibinfo{author}{\bibfnamefont{R.}~\bibnamefont{Mezzena}},
  \bibinfo{author}{\bibfnamefont{A.}~\bibnamefont{Ortolan}},
  \bibinfo{author}{\bibfnamefont{G.~A.} \bibnamefont{Prodi}},
  \bibinfo{author}{\bibfnamefont{L.}~\bibnamefont{Taffarello}},
  \bibnamefont{et~al.}, \bibinfo{journal}{Phys. Rev. Lett.}
  \textbf{\bibinfo{volume}{118}}, \bibinfo{pages}{021302}
  (\bibinfo{year}{2017}).

\bibitem[{\citenamefont{Budker and Romalis}(2007)}]{budker2007optical}
\bibinfo{author}{\bibfnamefont{D.}~\bibnamefont{Budker}} \bibnamefont{and}
  \bibinfo{author}{\bibfnamefont{M.}~\bibnamefont{Romalis}},
  \bibinfo{journal}{Nat. Phys.} \textbf{\bibinfo{volume}{3}}
  (\bibinfo{year}{2007}).

\bibitem[{\citenamefont{Weis et~al.}(2017)\citenamefont{Weis, Bison, and
  Gruji\'{c}}}]{Weis16}
\bibinfo{author}{\bibfnamefont{A.}~\bibnamefont{Weis}},
  \bibinfo{author}{\bibfnamefont{G.}~\bibnamefont{Bison}}, \bibnamefont{and}
  \bibinfo{author}{\bibfnamefont{Z.~D.} \bibnamefont{Gruji\'{c}}}, in
  \emph{\bibinfo{booktitle}{High Sensitivity Magnetometers}}, edited by
  \bibinfo{editor}{\bibfnamefont{A.}~\bibnamefont{Grosz}},
  \bibinfo{editor}{\bibfnamefont{M.~J.} \bibnamefont{Haji-Sheikh}},
  \bibnamefont{and} \bibinfo{editor}{\bibfnamefont{S.~C.}
  \bibnamefont{Mukhopadhyay}} (\bibinfo{publisher}{Springer},
  \bibinfo{year}{2017}), chap.~\bibinfo{chapter}{13}.

\bibitem[{\citenamefont{Alexandrov et~al.}(2002)\citenamefont{Alexandrov,
  Balabas, Budker, English, Kimball, Li, and Yashchuk}}]{Ale02}
\bibinfo{author}{\bibfnamefont{E.~B.} \bibnamefont{Alexandrov}},
  \bibinfo{author}{\bibfnamefont{M.~V.} \bibnamefont{Balabas}},
  \bibinfo{author}{\bibfnamefont{D.}~\bibnamefont{Budker}},
  \bibinfo{author}{\bibfnamefont{D.}~\bibnamefont{English}},
  \bibinfo{author}{\bibfnamefont{D.~F.} \bibnamefont{Kimball}},
  \bibinfo{author}{\bibfnamefont{C.-H.} \bibnamefont{Li}}, \bibnamefont{and}
  \bibinfo{author}{\bibfnamefont{V.~V.} \bibnamefont{Yashchuk}},
  \bibinfo{journal}{Phys. Rev. A} \textbf{\bibinfo{volume}{66}},
  \bibinfo{pages}{042903} (\bibinfo{year}{2002}).

\bibitem[{\citenamefont{Castagna et~al.}(2009)\citenamefont{Castagna, Bison,
  Di~Domenico, Hofer, Knowles, Macchione, Saudan, and Weis}}]{Cas09}
\bibinfo{author}{\bibfnamefont{N.}~\bibnamefont{Castagna}},
  \bibinfo{author}{\bibfnamefont{G.}~\bibnamefont{Bison}},
  \bibinfo{author}{\bibfnamefont{G.}~\bibnamefont{Di~Domenico}},
  \bibinfo{author}{\bibfnamefont{A.}~\bibnamefont{Hofer}},
  \bibinfo{author}{\bibfnamefont{P.}~\bibnamefont{Knowles}},
  \bibinfo{author}{\bibfnamefont{C.}~\bibnamefont{Macchione}},
  \bibinfo{author}{\bibfnamefont{H.}~\bibnamefont{Saudan}}, \bibnamefont{and}
  \bibinfo{author}{\bibfnamefont{A.}~\bibnamefont{Weis}},
  \bibinfo{journal}{Appl.~Phys.~B} \textbf{\bibinfo{volume}{96}},
  \bibinfo{pages}{763} (\bibinfo{year}{2009}).

\bibitem[{\citenamefont{Balabas et~al.}(2010)\citenamefont{Balabas, Karaulanov,
  Ledbetter, and Budker}}]{Bal10}
\bibinfo{author}{\bibfnamefont{M.~V.} \bibnamefont{Balabas}},
  \bibinfo{author}{\bibfnamefont{T.}~\bibnamefont{Karaulanov}},
  \bibinfo{author}{\bibfnamefont{M.~P.} \bibnamefont{Ledbetter}},
  \bibnamefont{and} \bibinfo{author}{\bibfnamefont{D.}~\bibnamefont{Budker}},
  \bibinfo{journal}{Phys. Rev. Lett.} \textbf{\bibinfo{volume}{105}},
  \bibinfo{pages}{070801} (\bibinfo{year}{2010}).

\bibitem[{ard()}]{ardui}
\emph{\bibinfo{title}{Arduino homepage}},
  \bibinfo{howpublished}{\url{http://www.arduino.cc}}, \bibinfo{note}{accessed:
  2017-11-08}.

\bibitem[{bos()}]{bosch}
\emph{\bibinfo{title}{Bosch {BNO}055}},
  \bibinfo{howpublished}{\url{https://www.bosch-sensortec.com/bst/products/all_products/bno055}},
  \bibinfo{note}{accessed: 2017-11-08}.

\bibitem[{hdf()}]{hdf5}
\emph{\bibinfo{title}{The {HDF} group homepage}},
  \bibinfo{howpublished}{\url{https://www.hdfgroup.org/}},
  \bibinfo{note}{accessed: 2017-11-08}.

\bibitem[{iee()}]{ieee754}
\bibinfo{note}{\url{https://en.wikipedia.org/wiki/IEEE_754}.}

\bibitem[{zli()}]{zlib}
\bibinfo{note}{{F}urthermore, the {HDF5} library is inherently combined with
  the zlib library, providing compression on demand, which enhances the
  efficient use of disk space, see {\url{https://zlib.net/}}.}

\bibitem[{\citenamefont{Kim}(1979)}]{Kim79}
\bibinfo{author}{\bibfnamefont{J.}~\bibnamefont{Kim}}, \bibinfo{journal}{Phys.
  Rev. Lett.} \textbf{\bibinfo{volume}{43}}, \bibinfo{pages}{103}
  (\bibinfo{year}{1979}).

\bibitem[{\citenamefont{Shifman et~al.}(1980)\citenamefont{Shifman, Vainshtein,
  and Zakharov}}]{Shi80}
\bibinfo{author}{\bibfnamefont{M.}~\bibnamefont{Shifman}},
  \bibinfo{author}{\bibfnamefont{A.}~\bibnamefont{Vainshtein}},
  \bibnamefont{and} \bibinfo{author}{\bibfnamefont{V.}~\bibnamefont{Zakharov}},
  \bibinfo{journal}{Nucl. Phys. B} \textbf{\bibinfo{volume}{166}},
  \bibinfo{pages}{493} (\bibinfo{year}{1980}).

\bibitem[{\citenamefont{Jackson~Kimball}(2015)}]{Kim15}
\bibinfo{author}{\bibfnamefont{D.~F.} \bibnamefont{Jackson~Kimball}},
  \bibinfo{journal}{New J. Phys.} \textbf{\bibinfo{volume}{17}},
  \bibinfo{pages}{073008} (\bibinfo{year}{2015}).

\bibitem[{\citenamefont{Budker et~al.}(2002)\citenamefont{Budker, Gawlik,
  Kimball, Rochester, Yashchuk, and Weis}}]{budker2002resonant}
\bibinfo{author}{\bibfnamefont{D.}~\bibnamefont{Budker}},
  \bibinfo{author}{\bibfnamefont{W.}~\bibnamefont{Gawlik}},
  \bibinfo{author}{\bibfnamefont{D.~F.} \bibnamefont{Kimball}},
  \bibinfo{author}{\bibfnamefont{S.~M.} \bibnamefont{Rochester}},
  \bibinfo{author}{\bibfnamefont{V.~V.} \bibnamefont{Yashchuk}},
  \bibnamefont{and} \bibinfo{author}{\bibfnamefont{A.}~\bibnamefont{Weis}},
  \bibinfo{journal}{Rev. Mod. Phys.} \textbf{\bibinfo{volume}{74}},
  \bibinfo{pages}{1153} (\bibinfo{year}{2002}).

\bibitem[{sci({\natexlab{a}})}]{scipy-spectrogram-note}
\bibinfo{note}{{E}ach spectrogram was created using the
  scipy.signal.spectrogram {P}ython module, see
  \url{https://docs.scipy.org/doc/scipy/reference/generated/scipy.signal.spectrogram.html}.}

\bibitem[{sci({\natexlab{b}})}]{scipy-welch-method-note}
\bibinfo{note}{{T}his procedure can be done using the scipy.signal.welch
  {P}ython module, see
  \url{https://docs.scipy.org/doc/scipy/reference/generated/scipy.signal.welch.html}.}

\bibitem[{\citenamefont{Cooley and Tukey}(1965)}]{Cooley1965}
\bibinfo{author}{\bibfnamefont{J.~W.} \bibnamefont{Cooley}} \bibnamefont{and}
  \bibinfo{author}{\bibfnamefont{J.~W.} \bibnamefont{Tukey}},
  \bibinfo{journal}{Math. Comput.} \textbf{\bibinfo{volume}{19}},
  \bibinfo{pages}{297} (\bibinfo{year}{1965}).

\bibitem[{\citenamefont{Allan}(1966)}]{Allan66}
\bibinfo{author}{\bibfnamefont{D.~W.} \bibnamefont{Allan}},
  \bibinfo{journal}{Proceedings of the IEEE} \textbf{\bibinfo{volume}{54}},
  \bibinfo{pages}{221} (\bibinfo{year}{1966}).

\bibitem[{\citenamefont{Yashchuk et~al.}(2000)\citenamefont{Yashchuk, Budker,
  and Davis}}]{yashchuk2000davll}
\bibinfo{author}{\bibfnamefont{V.~V.} \bibnamefont{Yashchuk}},
  \bibinfo{author}{\bibfnamefont{D.}~\bibnamefont{Budker}}, \bibnamefont{and}
  \bibinfo{author}{\bibfnamefont{J.~R.} \bibnamefont{Davis}},
  \bibinfo{journal}{Rev. Sci. Instrum.} \textbf{\bibinfo{volume}{71}},
  \bibinfo{pages}{341} (\bibinfo{year}{2000}).

\bibitem[{\citenamefont{Pustelny et~al.}(2016)\citenamefont{Pustelny, Schultze,
  Scholtes, and Budker}}]{Pustelny2016Dichroic}
\bibinfo{author}{\bibfnamefont{S.}~\bibnamefont{Pustelny}},
  \bibinfo{author}{\bibfnamefont{V.}~\bibnamefont{Schultze}},
  \bibinfo{author}{\bibfnamefont{T.}~\bibnamefont{Scholtes}}, \bibnamefont{and}
  \bibinfo{author}{\bibfnamefont{D.}~\bibnamefont{Budker}},
  \bibinfo{journal}{Rev. Sci. Instrum.} \textbf{\bibinfo{volume}{87}},
  \bibinfo{pages}{063107} (\bibinfo{year}{2016}).

\end{thebibliography}

\end{document}